\documentclass[aps,twocolumn,preprintnumbers,prd,superscriptaddress,nofootinbib,10pt,floatfix]{revtex4-2}
\usepackage[utf8]{inputenc} 
\usepackage{graphicx,graphics,xcolor,overpic,mathtools}
\usepackage{amsthm,amsmath,amssymb}
\usepackage[colorlinks]{hyperref}
\usepackage{textcomp}
\definecolor{coolblack}{rgb}{0.0, 0.18, 0.39}
\hypersetup{
    colorlinks = false,
   }
\PassOptionsToPackage{normalem}{ulem}

\usepackage{ulem} 
\usepackage{bm}
\usepackage{url}
\usepackage{physics}
\usepackage[makeroom]{cancel}
\usepackage{cleveref}
\usepackage{todonotes}
\usepackage{tensor}
\usepackage{mathrsfs}
\usepackage{slashed}
\usepackage[mode=buildnew]{standalone}

\newcommand{\comment}[1]{}

\usepackage{pgfplots}

\usepackage{diagbox}
\usepackage{stackengine,amssymb,graphicx}

\NewDocumentCommand{\evat}{sO{\bigg}mm}{%
  \IfBooleanTF{#1}
   {\mleft. #3 \mright|_{#4}}
   {#3#2|_{#4}}%
}
\usepackage{enumerate}
\definecolor{azure}{rgb}{0.0, 0.5, 1.0}

\usepackage{multirow}
\begin{document}

\title[]{Testing bosonic dark matter through white dwarf mass measurements}

\author{Jorge Castelo Mourelle}
\affiliation{
Departamento de Astronom\'{\i}a y Astrof\'{\i}sica, Universitat de Val\`encia,
Dr. Moliner 50, 46100 Burjassot (Val\`encia), Spain
}%
\affiliation{
Instituto de Ciencias Nucleares, Universidad Nacional Aut\'onoma de M\'exico, Circuito Exterior C.U., A.P. 70-543, M\'exico D.F. 04510, M\'exico.\\
}%

\author{Nicolas Sanchis-Gual}
\affiliation{
Departamento de Astronom\'{\i}a y Astrof\'{\i}sica, Universitat de Val\`encia,
Dr. Moliner 50, 46100 Burjassot (Val\`encia), Spain
}%

\author{Jos\'e A. Font}
\affiliation{
Departamento de Astronom\'{\i}a y Astrof\'{\i}sica, Universitat de Val\`encia,
Dr. Moliner 50, 46100 Burjassot (Val\`encia), Spain
}%
\affiliation{
Observatori Astronòmic, Universitat de València, C/ Catedrático José Beltrán 2, 46980, Paterna (València), Spain
}

\author{Juan Calder\'on Bustillo }
\affiliation{%
Departamento de F\'isica de Part\'iculas, Universidad de Santiago de Compostela and Instituto
Galego de F\'isica de Altas Enerxias (IGFAE), E-15782 Santiago de Compostela, Spain
}%
\affiliation{Department of Physics, The Chinese University of Hong Kong, Shatin, N.T., Hong Kong}

\date[ Date: ]{\today}
\begin{abstract}
Mass estimates of white dwarfs via electromagnetic methods, often differ from those obtained through gravitational redshift measurements, in some cases with  discrepancies ranging in $5$--$15\%$ across independent datasets.  
Although many of the discrepancies reported in large spectroscopic surveys and confirmed by high-precision techniques such as astrometric microlensing and wide-binary analyses may be attributable to thermal effects, model uncertainties or measurement errors prevent a complete description of some of the observations. Here, we explore an alternative explanation based on the presence of a gravitationally coupled bosonic scalar field that contributes to the stellar mass while remaining electromagnetically invisible. We construct stationary, static mixed configurations consisting of a white dwarf that presents a bosonic scalar field (dark matter) component, forming a composite white dwarf–boson star system. We explore families of solutions showing that a scalar field fraction of $f_{\rm DM} \sim 5$--$15\%$ to the mass contribution can account for the observed redshift excess. Our models provide a physically motivated explanation for the mass bias, might offer new observational signatures, and allow us to place preliminary constraints on the mass and compactness of the scalar field configuration. Finally, using our theoretical framework in combination with Bayesian model selection we provide plausible bounds for the mass of the constituent (ultralight) bosonic particle. 
\end{abstract}
\maketitle

\section{Introduction}

White dwarfs (WDs) are among the most common endpoints of stellar evolution, representing the final compact remnants of low- and intermediate-mass stars, in the range $\sim 0.8-8 M_{\odot}$ \cite{steen2024measuring}. Their internal structure, governed by electron degeneracy pressure, and their well-understood cooling sequences, make them excellent laboratories for astrophysical tests, including the study of gravitational theories and dark matter (DM) interactions. Because WDs are relatively long-lived and electromagnetically bright, they also provide precise observables such as surface gravity, effective temperature, photometric radius, and spectroscopic line profiles \cite{dufour2016montreal}.

Traditionally, the mass of a WD can be inferred through two independent methods: (i) electromagnetic analysis, which combines spectroscopy, photometry, and parallax to infer mass via model-dependent atmospheric parameters \cite{tremblay2009spectroscopic,calamida2022perfecting}; and (ii) gravitational redshift, which exploits the relativistic shift of spectral lines as a measure of the star’s compactness \cite{falcon2012gravitational}. Despite major improvements in data quality and theoretical models, occasional tensions between these approaches remain. While ensemble studies generally recover consistency with the canonical mass–radius relation, significant discrepancies arise in a non-negligible subset of objects, with typical offsets of a few percent and outliers reaching up to $\sim 15\%$ \cite{bergeron1995masses,raddi2025testing,arseneau2024measuring,chandra2020gravitational}. Explanations based on thermal effects, atmospheric modeling assumptions, or calibration uncertainties~\cite{arseneau2025resolution} can mitigate part of the tension but fall short of accounting for the full magnitude observed in the most extreme cases, motivating the exploration of alternative explanations. Large-scale surveys such as SDSS and Gaia-based analyses \cite{crumpler2024detection,chandra2020gravitational,steen2024measuring,de2023gaia,vallenari2023gaia}, together with high-precision probes including astrometric microlensing \cite{mcgill2023first}, broadly validate the theoretical relation, yet the persistent outliers motivate the consideration of additional physics as one possible explanation.

One intriguing possibility is that WDs might contain a gravitationally coupled but electromagnetically invisible component. In particular, scalar field DM, potentially forming a bosonic condensate within or around the star, can generate a gravitational mass defect that alters the observed compactness without affecting spectroscopic signatures. Such a component would naturally explain a mass discrepancy of order $\Delta M / M \sim 10\%$, mimicking a redshift excess without modifying the WD's atmospheric profile \cite{Sanchis-Gual:2022ooi,Brito:2015yfh,Leung:2013pra,Chan:2021gcm}. However, capture-based mechanisms of DM accumulation are severely constrained. For local DM densities ($\rho_{\mathrm{DM}}\sim 0.3\,\mathrm{GeV/cm^3}$) and direct detection bounds on cross-sections, the total DM mass acquired over a WD's lifetime is negligible, $f_{\mathrm{DM}} < 1\%$ \cite{Dasgupta:2019juq,Graham:2018efk}. To achieve fractions $f_{\mathrm{DM}} \sim 5\%$--$20\%$, a scalar field must have been present during the star's formation or gravitationally collapsed into a solitonic configuration prior to the WD phase \cite{Dasgupta:2019juq}. This demands highly non-standard scenarios that remain speculative but possible \cite{DiGiovanni:2020frc,Sanchis-Gual:2022ooi}. Other DM coupling options could be via modified gravity theories, in which the scalar field is described by pure non-minimal couplings of the gravitational field to the matter Lagrangian \cite{Jusufi:2025hte}.

Despite these limitations, the scalar field hypothesis remains compelling, especially in light of the broader context of exotic compact objects. Scalar boson stars (BSs), arising as solutions to the Einstein-(complex, massive) Klein-Gordon equations, are stable, horizonless, and transparent to light, making them viable DM candidates \cite{PhysRev.172.1331,Liebling:2012fv, CasteloMourelle:2025ujn}. Their mass range, spanning from planetary to supermassive and galactic scales~\cite{Schunck:2003kk,Mourelle:2025ilv} depending on the (ultralight) boson mass ($10^{-22}$ to $10^{-10}$ eV)~\cite{arvanitaki2010string,Freitas:2021cfi,Krasnov:2025aoo}, makes them astrophysically relevant \cite{Carvalho:2016jgc}. Moreover, mixed configurations involving both fermionic and bosonic matter, such as fermion-boson stars, have been shown to exhibit new stability branches and dynamical formation channels, particularly in neutron star contexts \cite{DiGiovanni:2020frc,DiGiovanni:2021vlu,Mourelle:2024qgo}.

In this work, we explore the scenario proposed in \cite{Sanchis-Gual:2022ooi}, namely the existence of a mixed WD–boson star system, constructed from equilibrium, in which the scalar field contributes a non-negligible fraction of the gravitational mass. The viability of equilibrium configurations involving a scalar field coupled to a fermionic star is strongly influenced by the relative length scales of the two components. Bosonic fields of different masses are associated with distinct characteristic spatial extensions, which determine their ability to form stable, coexisting configurations with a given stellar object. In particular, ultra-light bosons with galactic-scale wavelengths are incompatible with stellar-sized systems. In contrast, much heavier bosons, typically used to model neutron-star-scale objects, are confined to very compact regions and would form only small cores. Consequently, only an intermediate range of boson masses provides a suitable match to the characteristic size of WDs, allowing for consistent equilibrium configurations~\cite{Schunck:2003kk,Liebling:2012fv, CasteloMourelle:2025ujn}. 
Unlike neutron stars, WDs possess weaker gravitational fields and larger radii, making the resulting bosonic component to behave differently. Nevertheless, it could still significantly affect the gravitational redshift while leaving the atmospheric modeling practically unchanged. We find that our model not only provides an explanation for the observed mass biases between electromagnetic and gravitational redshift methods but also opens a new observational window into the potential interplay between luminous compact stars and DM. Future observational strategies to test this proposal are suggested, including searches for environmental correlations (e.g.~denser DM regions like globular clusters \cite{Ramirez-Quezada:2022uou}) and asteroseismological signatures.

The structure of the paper is as follows: In \Cref{section2}, we introduce the theoretical framework for mixed WDs used in our model. \Cref{section3} describes the numerical algorithm to obtain the equilibrium configurations, together with some key physical magnitudes.  In \Cref{section4} we present the observational data employed in the analysis and discuss the discrepancies between different measurement methods applied to the same astrophysical objects. \Cref{section5} is devoted to comparing the equilibrium models, generated using various parameterizations, against the observational dataset. Next, in \Cref{section7} we perform a quantitative Bayesian analysis assessing the agreement between our numerical simulations of WD for different boson masses and observational data. In particular, we first perform parameter inference on the underlying boson mass responsible for the scalar component of the mixed WDs. Second, we compare the statistical preference for such mixed WD model with respect to a model assuming equal gravitational and electromagnetic mass measurements, i.e., a pure WD omitting any scalar field component. This is followed by a discussion of astrophysical scenarios in which this approach may be particularly relevant, together with final remarks and conclusions in \Cref{section8}. The paper closes with two appendices. \Cref{AppendixA} discusses technical details on the physical scales involved and the derivation of observable quantities, while \Cref{AppendixB} is devoted to error propagation calculations.

\section{Theoretical setup }
\label{section2}

The equilibrium configurations for our mixed systems are defined by the coupled Einstein–Klein–Gordon system together with the general relativistic hydrodynamics equations
\begin{equation}
    R_{\alpha\beta}-\frac{1}{2}Rg_{\alpha\beta}=G_{\alpha\beta}=8\pi T_{\alpha\beta}.
\end{equation}
Here, $R_{\alpha\beta}$ is the Ricci tensor of the 4-dimensional spacetime, $g_{\alpha\beta}$ is the spacetime metric, $R$ is the Ricci scalar, and $T_{\alpha\beta}$ represents the stress-energy tensor of the matter content. We use units in which $G=c=\hbar=1$.

Since we aim to model WDs we require a stress-energy tensor that accounts for their fermionic nature. To this end, we adopt the perfect fluid approximation, which allows us to describe the fermionic matter source as:
\begin{equation}
    T^{\rm F}_{\alpha\beta}=\left( \rho+p\right)u_{\alpha}u_{\beta}+pg_{\alpha\beta},
\end{equation}
where $\rho$ is the  energy density and $p$ is the fermionic pressure. The fourth-velocity is given by $u^{\alpha}=(-1/\sqrt{-g_{tt}},0,0,0)$.

To describe the thermodynamics of the fermionic part of the mixed WD we assume a polytropic law,
\begin{equation}
    p=K\rho^{\Gamma},
\label{politrop}
\end{equation}
where $\Gamma=1+1/n$ is the adiabatic index and $K$ is the polytropic constant, which plays the role of a scaling factor. Since $K^{n/2}$ has units of length, we can use it to redefine dimensionless quantities and fix the scales. This is a well-known approach for building WD models \cite{Sanchis-Gual:2022ooi}. In our case, the polytropic index for average low-mass WDs is set to  $N = 1.5$ (i.e.~$\Gamma = \frac{5}{3}$) and we use a value of $ K \sim 5000$ to recover the desired scales. However, it is essential to note that $K$ is rescaled from the system of equations, and we restore its original value after the numerical calculations.

Although the polytropic approximation is well-motivated and widely used, it remains an approximation. This imposes certain limitations on the accuracy of our results. As we will see, our models reproduce the observed data reasonably well in the low-mass regime. However, for more compact configurations, it becomes evident that a more refined treatment of the thermodynamics would be necessary, either by exploring different values of the polytropic index $n$ and constant $K$, or by adopting realistic equations of state that account for temperature and additional microphysical effects. Nevertheless, since our goal is not to perform a detailed analysis of the fermionic component itself, but rather to investigate the impact of the scalar field on the total mass and structure of composite WDs, we regard the polytropic approximation as sufficient for our purposes.

In our approach we have a second matter component sourcing a complex scalar field, which contributes to the total mass. The canonical stress-energy tensor of the scalar field reads,
\begin{equation}
T^{\rm SF}_{\alpha\beta}=2\nabla_{(\alpha}\Phi^*\nabla_{\beta)}\Phi-g_{\alpha\beta}\left[\nabla^{\mu}\Phi^*\nabla_{\mu}\Phi+V\left(|\Phi|^2\right)\right]],
    \label{stressBS}
\end{equation}
where $\Phi$ is the scalar field (with $\Phi^*$ its complex conjugate) and $V(|\Phi|^2)$ is the potential for the scalar field, which in this work is given by $V(|\Phi|^2)=\mu^2|\Phi|^2$. We use this potential because in the configurations considered here the bosonic field amplitudes are small, which strongly suppresses the contribution of self-interactions. We explicitly verified this by including quartic terms of the form $\pm \Lambda |\Phi|^4$ and varying the self-coupling constant over a wide range, finding no significant changes in the resulting equilibrium solutions, even for extreme values of $\Lambda$. This behavior is expected, since near the minimum of the potential self-interaction terms enter only as higher-order corrections, rendering their impact on the structure of the solutions negligible within the parameter space explored. A clear illustration of this effect in the rotating case can be found in Fig.~1 of~\cite{Adam:2022nlq}.

The conserved current linked to the $U(1)$ symmetry of the Lagrangian plays a crucial role in the stability of the field. According to Noether's theorem, the current related to the scalar field stress-energy tensor is given by
\begin{equation}
    j^{\alpha}=\frac{i}{2}g^{\alpha\beta}\left(\Phi^*\nabla_{\beta}\Phi-\Phi\nabla_{\beta}\Phi^*\right)
\end{equation}
and satisfies the following conservation law,
\begin{equation}
    \nabla_{\alpha}j^{\alpha}=\frac{1}{\sqrt{-g}}\partial_{\alpha}\left(\sqrt{-g} \, j^{\alpha}\right)\,,
\end{equation}
with $g=\det(g_{\alpha\beta})$.
Our two-component matter system is minimally coupled to gravity, such that both matter sources only interact  gravitationally. The equations governing the equilibrium of the system are given by:
\begin{equation}
    \begin{split}
        &G_{\alpha\beta}=8\pi(T^{\rm F}_{\alpha\beta}+T^{\rm SF}_{\alpha\beta}),\\
        &g^{\alpha\beta}\nabla_\alpha\nabla_{\beta}\Phi=\frac{dV}{d|\Phi|^2}\Phi,\\
        &\nabla_{\alpha}T_f^{\alpha\beta}=0.
    \end{split}
    \label{FBSEQS}
\end{equation}
The above set of equations describes a system in which a polytropic fermionic source and a bosonic scalar field are coupled through gravity to form a compact object. However, we have not yet imposed any assumptions regarding the geometry or symmetries of the system. Consequently, the choice of the spacetime metric will depend on the kind of solution we look for. In this study, we will treat non-spinning static solutions.

It is worth noting that several studies have been conducted on the static aspects of mixed fermion boson stars, specifically treating the fermionic part as a neutron star~\cite{DiGiovanni:2020frc,Nyhan:2022pda,DiGiovanni:2022mkn,DiGiovanni:2021vlu,PhysRevD.87.084040,Kain:2021bwd}. Moreover, in \cite{Sanchis-Gual:2022ooi} fermion-boson systems have been dynamically explored under the WD scope, as mentioned before.
To solve a general, static, spherically symmetric system, we will use Schwarzschild-like coordinates, yielding the following metric element \cite{DiGiovanni:2020frc},
\begin{equation}
    ds^2=-b^2dt^2+a^2dr^2+(r^2d\theta+r^2\sin^2\theta d\psi^2) ,
    \label{metricbs}
\end{equation}
where $b$ and $a$ are the metric functions and depend only on the radial coordinate $r$.
For the bosonic part, we need to close the problem with a given ansatz for the scalar field.  To meet the symmetry requirements, the scalar field has the form, 
\begin{equation}
    \Phi(t,r)=\phi_s(r)e^{-i w_st},
    \label{fieldansatz1}
\end{equation}
where $ w_s$ is the fundamental oscillation frequency of the scalar field. The modified Tolman-Oppenheimer-Volkoff (TOV) system is obtained through Eq.~(\ref{FBSEQS}) by using Eqs.~(\ref{metricbs}) and (\ref{fieldansatz1}). This leads to the equilibrium equations for the mixed WD with a coupled scalar field,
\begin{equation}
    \begin{split}
        &\frac{db}{dr}=\frac{b}{2}\left(\frac{a^2-1}{r}+8\pi r\left[\left(\frac{w_s^2}{b^2}-\mu^2\right)a^2\phi_s^2+\Psi^2+2a^2p\right]\right),\\
         &\frac{da}{dr}=\frac{a}{2}\left(\frac{1-a^2}{r}+8\pi r\left[\left(\frac{w_s^2}{b^2}+\mu^2\right)a^2\phi_s^2+\Psi^2+2a^2\rho\right]\right),\\
         &\frac{d\phi_s}{dr}=\Psi,\\
         &\frac{d\Psi}{dr}=-\left(1+a^2-8\pi r a^2\left[\mu^2\phi_s^2+\frac{1}{2}(\rho-p)\right]\right)\frac{\Psi}{r}\\
         &\quad\quad\,\,\,\,
         -\left(\frac{w_s^2}{b^2}-\mu^2\right)a^2\phi_s,\\
         &\frac{dp}{dr}=-(\rho+p)\frac{1}{a}\frac{db}{dr},
    \end{split}\label{syseq}
\end{equation}
where $\mu$ is the mass of the bosonic particle. This system is formally consistent with the modified TOV equations commonly employed in studies of mixed neutron–boson stars \cite{HENRIQUES198999,Jetzer:1991jr,Pisano:1995yk,deSousa:1995ye,DiGiovanni:2020frc,Mourelle:2024qgo,Lazarte:2025etl}. It should be emphasized, however, that in the present work the polytropic equation of state (EOS) we employ is specifically tailored to describe a WD-type fluid, as discussed previously.
From the above system of equations, the gravitational total mass of the mixed star is computed as
\begin{equation}
M^{\rm T}=\lim_{r\rightarrow\infty}\frac{r}{2}\left(1-\frac{1}{a^2}\right).
\label{admmass}
\end{equation}
The resulting modified TOV system describes mixed configurations and naturally reduces to known limits: a pure boson star when $ p = \rho = 0 $, and a pure polytropic WD when the scalar field vanishes, $\phi_s = 0$.

\section{Numerical approach}
\label{section3}

\subsection{Method}


The system of equations (\ref{syseq}) forms an eigenvalue problem for the scalar field frequency  $w_s$, which is solved using a shooting method. For a given central value $\phi_0 = \phi_s(r=0) $, there exists a critical shooting frequency $ w_{\text{shot}} $ that ensures regularity and decay of the scalar field at large radii. The differential equations are integrated using a fourth-order Runge-Kutta (RK4) method.

Once the correct eigenvalue is found, the metric component $g^{tt}$  and the frequency must be rescaled by their asymptotic values, specifically  $g^{tt}(r \to \infty)$, to recover physical observables. In the absence of a scalar field (i.e., in the pure WD case), the system no longer requires shooting, and equilibrium can be directly obtained by standard RK integration.

To ensure a correct interpretation of the output profiles for each model (mass, radius, density, field amplitude, etc.), it is essential to clearly define the units and characteristic scales used. Those are discussed in detail in \Cref{AppendixA}.

\subsection{Komar mass}

Although Eq.~(\ref{admmass}) allows us to determine the total gravitational mass of the system, we are particularly interested in quantifying the individual contributions from the bosonic scalar field and the fermionic component. This will enable us to evaluate the fraction of the total mass of the star attributable to the scalar field.

We obtain the WD and the scalar field contributions to the mass by evaluating the Komar integral~\cite{stergioulas1994comparing}. For stationary spacetimes with a temporal Killing vector  $\xi^\mu$, the Komar mass is generally defined as
\begin{equation} \label{eq:KomarGeneral}
M_{\rm Komar} = -\frac{1}{8\pi} \int_S \epsilon_{\mu\nu\rho\sigma}\, \nabla^\rho \xi^\sigma \, dS^{\mu\nu}\,,
\end{equation}
where symbol \(\epsilon_{\mu\nu\rho\sigma}\) is the (four–dimensional) Levi–Civita tensor, totally antisymmetric, and $dS^{\mu\nu}$ is the oriented surface element bivector of the two-dimensional surface $S$. An equivalent expression for the Komar mass is given by
\begin{equation} \label{eq:KomarGeneral2}
M_{\rm Komar} = 2 \int_\Sigma \left( T_{\mu\nu} - \frac{1}{2}T\, g_{\mu\nu}\right) n^\mu \xi^\nu \, dV\,,
\end{equation}
where  $T = g^{\mu\nu} T_{\mu\nu}$ is the trace of the stress-energy tensor, $n^\mu$ is a timelike unit vector normal to the spatial hypersurface $\Sigma$, and $dV$ is the volume element at $\Sigma$.

For the pure WD part the Komar mass reads~\cite{PhysRev.113.934,Abreu:2010sc},
\begin{equation} \label{eq:KomarNSmatter}
M_{\rm Komar}^{\rm WD} = 4\pi \int_0^R dr\, r^2\, \frac{b(r)}{a(r)} \Bigl[\rho(r)+3p(r)\Bigr]\,,
\end{equation}
where $R$ is the star's radius.  
Similarly, for the scalar field part, the Komar mass integral takes the following form~\cite{PhysRev.187.1767,Jetzer:1991jr}:
\begin{equation} \label{eq:KomarBS}
M_{\rm Komar}^{\rm BS} = 8\pi \int_0^R dr\, a(r)\, r^2 \left[\frac{2w_s^2\, \phi_s(r)^2}{b(r)} - \frac{b(r)}{2}\, V\Bigl(|\phi_s(r)|^2\Bigr)\right]\,.
\end{equation}

The total mass is given again by the addition of both terms,
\begin{equation}
    M_{\rm Komar}^{\rm T}=M_{\rm Komar}^{\rm WD}+M_{\rm Komar}^{\rm BS},
\end{equation}
that, up to a certain error due to numerical uncertainties, coincides with the ADM mass~\ref{admmass}. 

\section{Observational data}
\label{section4}

As mentioned before, WD masses are estimated from astrophysical observations by two independent methods: {(i)} electromagnetic emission (EM) and { (ii)} gravitational redshift (GRS). Here we briefly discuss the two approaches and how the difference
$\Delta M = M_{\rm GRS} - M_{\rm EM}$ is defined. This is of key importance in our study, since we associate the quantity $\Delta M$ with $f_{\rm DM}$, the fraction of DM present in our composite stars. We defer to \Cref{AppendixB} the discussion of the computation of the error propagation associated with the mass difference. We do not delve into full technical details of the measurement methodologies. Instead, our objective is to provide a general overview of the fundamental principles behind the two main observational approaches, and to clarify how these can inform and motivate the scenario we propose.

\subsection{Electromagnetic method (EM)}
\label{sec:metodoEM}

The EM method combines high-resolution spectroscopy, primarily of Balmer lines for DA WDs, or  He\,\textsc{i} lines for DB/DBA types, with broad-band photometry and precise parallax measurements from astrometric missions such as {Gaia} \cite{tremblay2020gaia}. This multi-step procedure allows for a model-dependent determination of the WD's mass, $M_{\rm EM}$, via the following steps:

\begin{enumerate}
  \item The observed absorption-line profiles are fitted to grids of NLTE (non-local thermodynamic equilibrium) atmosphere models \cite{bohlin2020new}, yielding estimates of the effective temperature $T_{\rm eff}$ and the surface gravity $\log \tilde g$, typically with uncertainties on the order of $\sim 1\%$ \cite{koester2010white}. 
  
  \item The surface gravity $\tilde g$ is defined by
  \begin{equation}
    \tilde g = \frac{G\,M_{\rm EM}}{R_s^2}
    \quad\Longrightarrow\quad
    M_{\rm EM} = \frac{\tilde g\,R_s^2}{G},
  \end{equation}
  where $R_s$ is the stellar radius and $G$ is the gravitational constant. Hence, the determination of $M_{\rm EM}$ requires an independent estimate of $R_s$.
  
  \item The radius $R_s$ is inferred from the Stefan–Boltzmann law using the effective temperature and the bolometric luminosity:
  \begin{equation}
    L = 4\pi R_s^2\,\sigma\,T_{\rm eff}^4,
  \end{equation}
  where $L$ is computed from the observed flux (via broad-band photometry) and the parallax-based distance, and $\sigma$ is the Stefan–Boltzmann constant \cite{tremblay2011improved}.
\end{enumerate}

This approach yields a model-dependent mass estimate that relies on atmospheric models and precise photometric and astrometric data. While generally robust, it may be affected by systematic uncertainties in the atmospheric modeling and assumptions about the WD’s composition.

\subsection{Gravitational redshift method (GRS)}
\label{sec:metodoGR}

According to general relativity, photons escaping from the gravitational potential well of a WD undergo gravitational redshift. This effect manifests as a systematic shift in the observed spectral lines. After correcting the total line shift $\Delta v$ for the star’s radial motion, one isolates the component due purely to gravity, denoted $v_{\rm GRS}$ \cite{falcon2010gravitational}. This quantity provides a model-independent estimate of the WD mass via the relation:
\begin{equation}
    v_{\rm GRS} = \frac{G\,M_{\rm GRS}}{c\,R_s}
  \quad\Longrightarrow\quad
  M_{\rm GRS} = \frac{v_{\rm GRS}\,c\,R_s}{G},
\end{equation}
where $M_{\rm GRS}$ is the gravitational mass 
and $c$ is the speed of light. 

For practical applications, this relation is often expressed in convenient astrophysical units:
\begin{equation}
    v_{\rm GRS}\,[\mathrm{km/s}] = 0.6365\,\frac{M_{\rm GRS}/M_\odot}{R_s/R_\odot}\,,
\end{equation}
which directly links the gravitational redshift (in km/s) to the mass and radius expressed in solar units. Since this method depends only on the geometry of spacetime and the WD’s radius, often derived from parallax and photometry, it serves as a powerful, largely model-independent probe of the stellar mass.

\subsection{Mass difference $\Delta M$}

To quantify the observational bias between the two methods we define
 $  \Delta M = M_{\rm GRS} - M_{\rm EM}$.
We argue that this quantity can be directly linked with the fraction of DM in the composite stars, $f_{\rm DM}$, as it represents the discrepancy between the gravitational and luminous mass measurements. 

\begin{figure}[t!]
\includegraphics[clip,width=1.0\columnwidth]{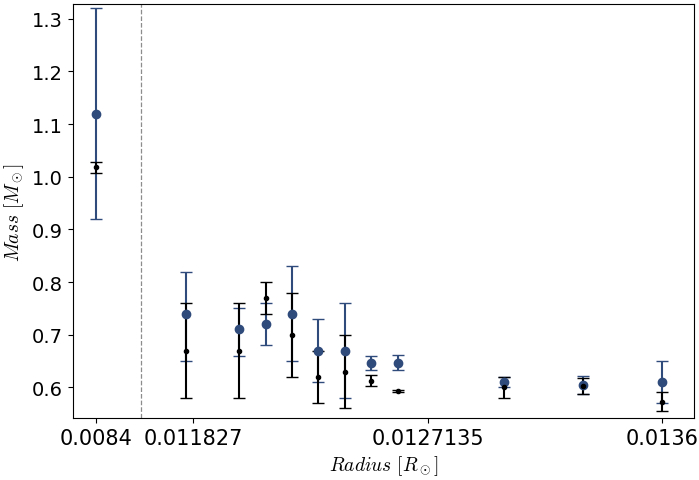}
\caption{Sample of WDs for which pairs of observational mass measurements are available. Small black circles correspond to the EM measurement, while big blue circles indicate the GRS counterpart. Each measurement is accompanied by its corresponding uncertainty.  In ascending order of radii, the stars reported are: Sirius B \cite{joyce2018gravitational,bond2017sirius}, 
DB  \cite{falcon2012gravitational,2011ApJ...737...28B}, 
DBA  \cite{falcon2012gravitational,2011ApJ...737...28B}, 
Hyades WD \cite{2019A&A...627L...8P,tremblay2012spectroscopic}, 
DBs  \cite{falcon2012gravitational,2011ApJ...743..138G}, 
Chandra \cite{2020ApJ...899..146C,arseneau2024measuring}, 
El-Badry \cite{arseneau2024measuring,arseneau2024measuring}, 
DA \cite{2010ApJ...712..585F,2011ApJ...730..128T}, 
DAs  \cite{2010ApJ...712..585F,2011ApJ...730..128T}, 
Koester  \cite{koester1987gravitational,2011ApJ...743..138G},
Procyon B \cite{Onofrio:2014txa,2000AJ....119.2428G}  and  
40 Eridani B \cite{popper1954red,bond2017sirius}. The $x$-axis is broken in the value indicated by the vertical dotted line to facilitate the visualisation of the data points.}
\label{stars_obs}
\end{figure}

Fig.~\ref{stars_obs} displays a set of WD masses measured using both methods. The models were selected on the basis of having both types of measurements available from independent determinations. It is important to note that some of the observational values displayed in Fig.~\ref{stars_obs}, although associated with stars, do not correspond to individual physical objects. Instead, they represent average magnitudes obtained from groups of stars with similar characteristics, as computed in the corresponding references (indicated in the caption). This is the case for the DB, DBs, DA, DAs, and DBA types in Fig.~\ref{stars_obs}.
At first glance, the two mass measurements are in close agreement for objects with lower masses and larger radii, while the opposite holds for more massive and compact stars. This general trend could imply that the latter stars could host a higher amount of scalar field in the form of DM and be more prone to DM accretion. Conversely, for less massive and more diffuse stars, the discrepancy between EM and GRS mass estimates becomes almost negligible. This suggests that in such systems, the scalar field is either weakly supported or entirely absent. Whether these trends hold consistently at the percentage level will be assessed in the following numerical analysis. 

\section{Equilibrium stellar models}
\label{section5}


\begin{table*}[t!]
\centering
\caption{Equilibrium stellar models for four of the observed WDs in our sample, each for three different scalar field configurations. The first subtable corresponds to a boson mass of $\mu=1\times 10^{-10}\,\rm eV$, which leads to a scalar field profile concentrated in a compact central core, effectively forming an inner DM component. The second configuration, with $\mu=5.04\times 10^{-11}\,\rm eV$, produces models where the scalar field occupies an intermediate internal region, distinct from both the core and outer layers. Finally, for $\mu=1.68\times 10^{-11}\,\rm eV$, the scalar field extends over a region comparable in size to that of the fermionic matter, resulting in a more diffuse distribution.}
\begin{tabular}{c|cccc|ccccc|cc}

\hline
\textbf{WD}& $M_{\rm GRS}$ & $M_{\rm EM}$ & $R$ & $\Delta M$ & $M_{\rm tot}$ & $M_{\rm WD}$  & $M_{\rm SF}$& $\% M_{\rm bosonic}$  & $R$ &  $p_{0}$& $\phi_{0}$ \\
 & $(\mathrm{M_{\odot}})$ 
 & $(\mathrm{M_{\odot}})$ 
 & $(R_{\odot})$
 &  $(\mathrm{M_{\odot}})$ 
 & $(\mathrm{M_{\odot}})$
 & $(\mathrm{M_{\odot}})$
 & $(\mathrm{M_{\odot}})$
 & 
 &$\rm (km)$
 & (code units)
 & (code units)\\
\hline
\multicolumn{12}{c}{$\mu=1\times 10^{-10}$ eV} \\
\hline
Sirius B \cite{joyce2018gravitational} & $1.12$ & $1.018$ & $0.0084$ & $0.102$   &$1.12$ & $1.018$ & $0.101$ & $\sim 9\%$ & $\sim 6300$ & $47.5\times10^{-10}$ & $1.03\times 10^{-4}$ \\
 DB \cite{falcon2012gravitational} & $0.74$ & $0.67$ & $0.0118$ & $0.07$  & $0.75$ & $0.69$ & $0.059$ & $\sim 7\%$ & $\sim 7800$ & $5.02\times 10 ^{-10}$ & $3.60\times 10^{-5}$ \\
 DBA \cite{falcon2012gravitational} & $0.71$ & $0.67$ & $0.012$ & $0.04$  & $0.72$ & $0.65$ & $0.06$ & $\sim6\%$ & $\sim7850$ & $5.01\times10^{-10}$ & $3.72\times 10^{-5}$ \\
Chandra  \cite{chandra202199} & $0.67$ & $0.62$ & $0.0123$ & $0.05$  & $0.7$ & $0.64$ & $0.058$ & $\sim 7\%$ & $\sim7900$ & $5.001\times10^{-10}$ & $3.52\times 10^{-5}$ \\
\hline
\multicolumn{12}{c}{$\mu = 5.05\times 10^{-11}$ eV} \\
\hline
Sirius B \cite{joyce2018gravitational} & $1.12$ & $1.018$ & $0.0084$ & $0.102$   &$1.11$ & $1.017$ & $0.094$ & $\sim 8\%$ & $\sim 7000$ & $6.15\times10^{-10}$ & $2.79\times 10^{-5}$ \\
 DB \cite{falcon2012gravitational} & $0.74$ & $0.67$ & $0.0118$ & $0.07$  & $0.747$ & $0.674$ & $0.0747$ & $\sim 10\%$ & $\sim 7900$ & $1.80\times 10 ^{-10}$ & $1.80\times 10^{-5}$ \\
 DBA \cite{falcon2012gravitational} & $0.71$ & $0.67$ & $0.012$ & $0.04$  & $0.71$ & $0.66$ & $0.05$ & $\sim7\%$ & $\sim8200$ & $0.95\times10^{-10}$ & $1.10\times 10^{-5}$ \\
Chandra  \cite{chandra202199} & $0.67$ & $0.62$ & $0.0123$ & $0.05$  & $0.67$ & $0.62$ & $0.059$ & $\sim 8\%$ & $\sim8300$ & $1.00\times10^{-10}$ & $1.30\times 10^{-5}$ \\
\hline
\multicolumn{12}{c}{$\mu = 1.68\times 10^{-11}$ eV } \\
\hline
Sirius B \cite{joyce2018gravitational} & $1.12$ & $1.018$ & $0.0084$ & $0.102$   &$1.12$ & $1.02$ & $0.103$ & $\sim 9\%$ & $\sim 7000$ & $2.75\times10^{-10}$ & $1.8\times 10^{-5}$ \\
 DB \cite{falcon2012gravitational} & $0.74$ & $0.67$ & $0.0118$ & $0.07$  & $0.739$ & $0.670$ & $0.067$ & $\sim 9\%$ & $\sim 8100$ & $0.62\times 10 ^{-10}$ & $1.0\times 10^{-5}$ \\
 DBA \cite{falcon2012gravitational} & $0.71$ & $0.67$ & $0.012$ & $0.04$  & $0.71$ & $0.67$ & $0.04$ & $\sim5\%$ & $\sim8500$ & $0.45\times10^{-10}$ & $0.7\times 10^{-5}$ \\
Chandra  \cite{chandra202199} & $0.67$ & $0.62$ & $0.0123$ & $0.05$  & $0.669$ & $0.615$ & $0.053$ & $\sim 8\%$ & $\sim8500$ & $0.41\times10^{-10}$ & $0.8\times 10^{-5}$ \\
\hline
\end{tabular}
\label{table1}
\end{table*}

  
We explore equilibrium configurations of WDs coupled to a scalar field motivated by the observed $\sim 5\%$ to $12\%$ discrepancies between mass estimates derived from GRS and EM methods (cf.~Fig.~\ref{stars_obs}). To address this, we adopt a dual approach that combines observationally guided modeling with the theoretical construction of mixed fermion–boson stellar systems. For several of the observed WDs with available mass and radius estimates, we search for initial conditions that yield equilibrium solutions consistent with the observational data. To do so we focus, specifically, on the values of the central pressure and scalar field amplitude. This procedure allows us to construct a set of individual models whose total mass, WD and bosonic mass components, and approximate radii agree remarkably well with the measured values. Although the fermionic matter is modeled using a simplistic polytropic EOS, the resulting configurations maintain a high degree of consistency with observations. The location and extent of the scalar field are crucial, as they influence the behavior of the fermionic component of the stellar models and affect the overall matter distribution and structure of the equilibrium configurations, as we will analyze in the sections that follow.

\subsection{Representative models}

Table~\ref{table1} reports the class of representative (or fiducial) models of our large sample. The last two columns include the corresponding central pressure and scalar field amplitude used to generate each configuration. For each observational case we explore solutions within three different theoretical frameworks, corresponding to distinct scalar field configurations. Specifically, we fix the scalar boson mass to the following  three values: $\mu = (10, 5.04, 1.68) \times 10^{-11}\,\mathrm{eV}$. These values are not arbitrary. The chosen order of magnitude for $\mu$ is physically well motivated in the context of composite WDs as it yields scalar field configurations with spatial extents of several hundred to a few thousand kilometers and mass contributions ranging from a few percent up to nearly one solar mass. In this regime, the scalar component naturally operates on the same characteristic scales as WDs. Note that a bosonic particle with a mass far outside this range would not generate configurations with observable consequences for WDs, as the mass of the scalar field is directly related to the spatial extent of its distribution~\cite{Schunck:2003kk,Liebling:2012fv}, and the scales would be incompatible. 
Very small boson masses ($\mu\sim10^{-22}-10^{-20}$ eV) correspond to Compton wavelengths of galactic scales \cite{Schunck:1998nq,Mourelle:2024dlt,Mourelle:2025ilv}, whereas larger masses lead to scalar-field configurations confined to regions of only a few kilometers, comparable to the size of a neutron star or even smaller, where the field effectively acts as a compact dark-matter core \cite{Lai:2004fw,DiGiovanni:2020frc,Mourelle:2024qgo}.

On the other hand, the specific choice of $\mu$ allows us to investigate how variations in boson mass affect the spatial structure of the scalar field. Even small changes in $\mu$ lead to qualitatively different configurations: for $\mu = 1 \times 10^{-10}\,\mathrm{eV}$, the scalar field forms a compact central core, while for $\mu = 1.68 \times 10^{-11}\rm eV$, it spreads over a volume comparable to that of the fermionic component~\cite{ brito2015accretion,Brito:2015yfh}. Therefore, the three selected values probe a range of mixed configurations in which the scalar field plays distinct structural roles.

\begin{figure*}
    \centering
    \includegraphics[width=0.32\textwidth]{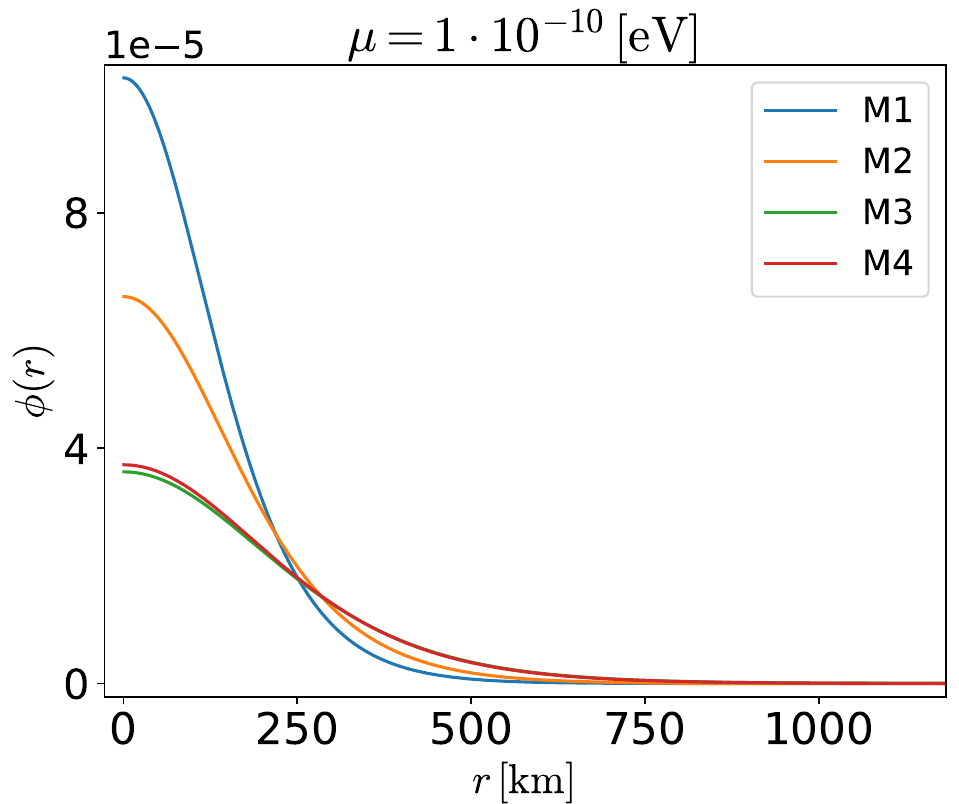}\hfill
    \includegraphics[width=0.32\textwidth]{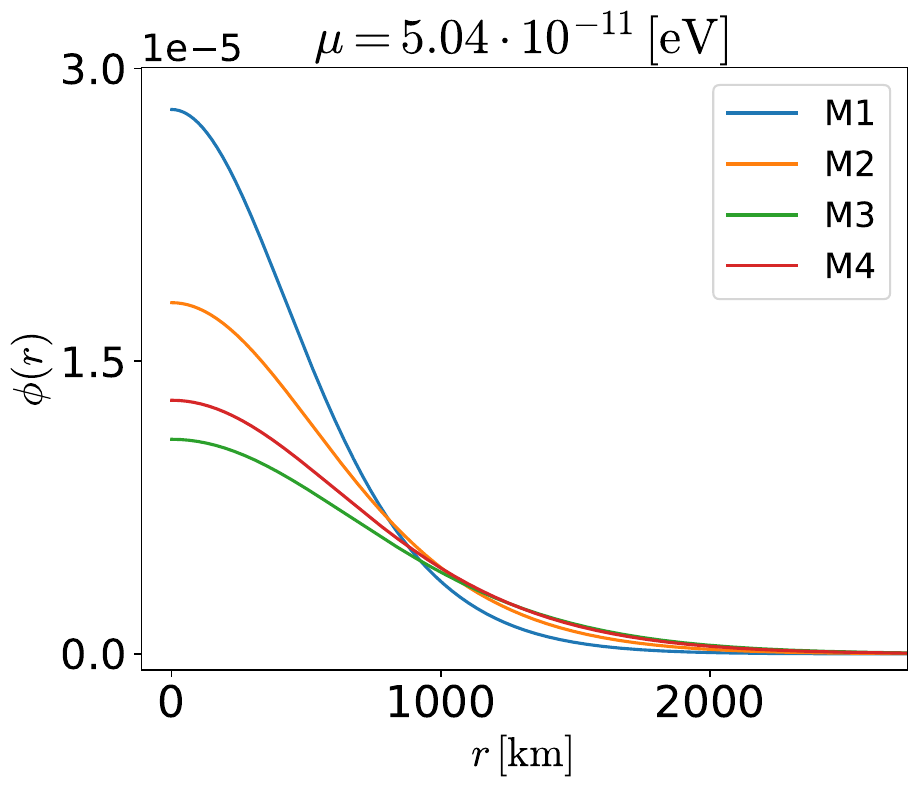}\hfill    
    \medskip
    \includegraphics[width=0.32\textwidth]{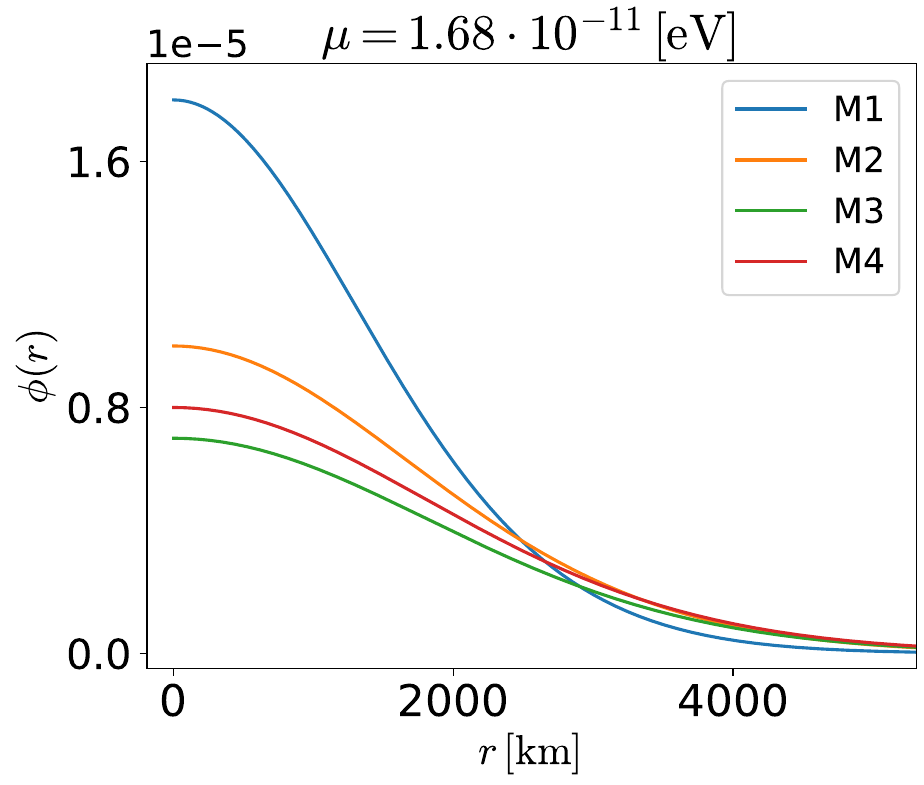}\hfill
    \includegraphics[width=0.32\textwidth]{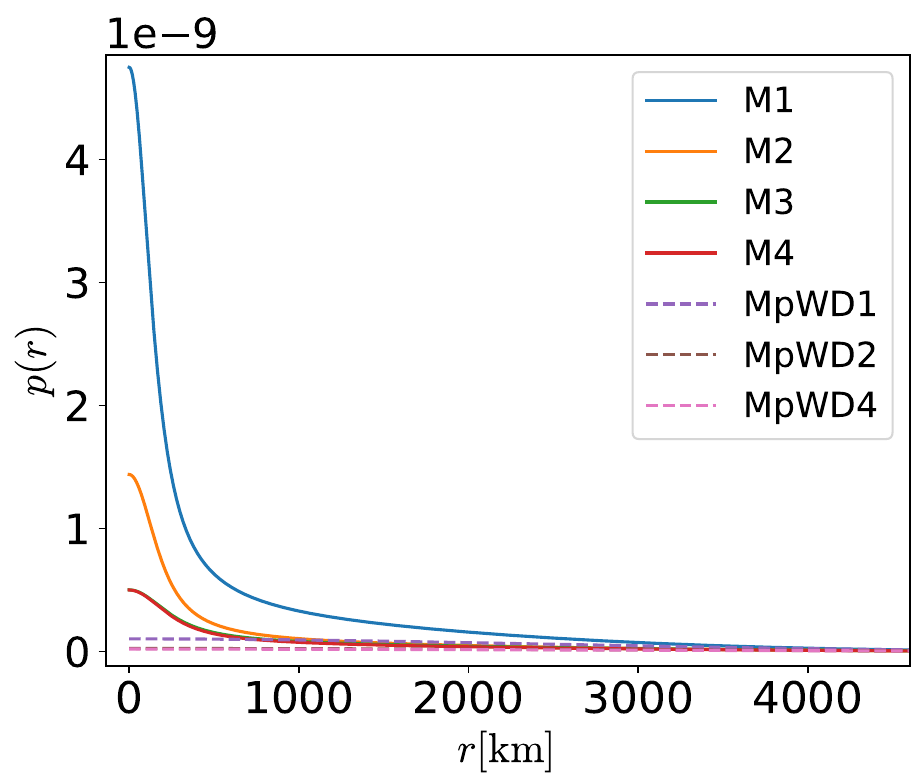}\hfill    
    \medskip
    \includegraphics[width=0.32\textwidth]{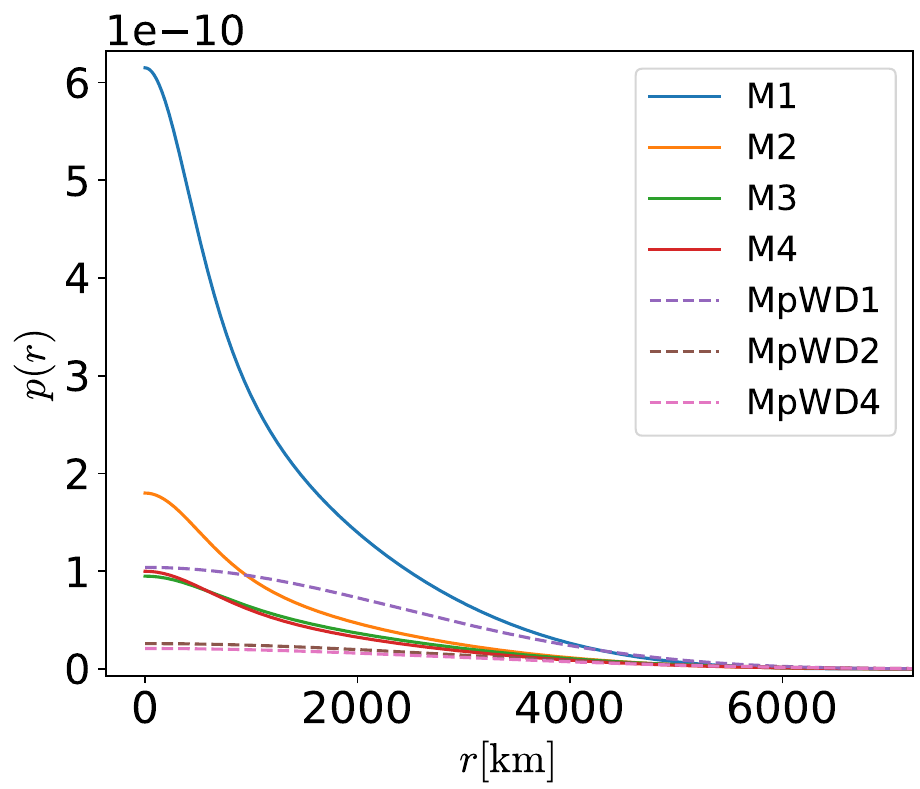}\hfill
    \includegraphics[width=0.32\textwidth]{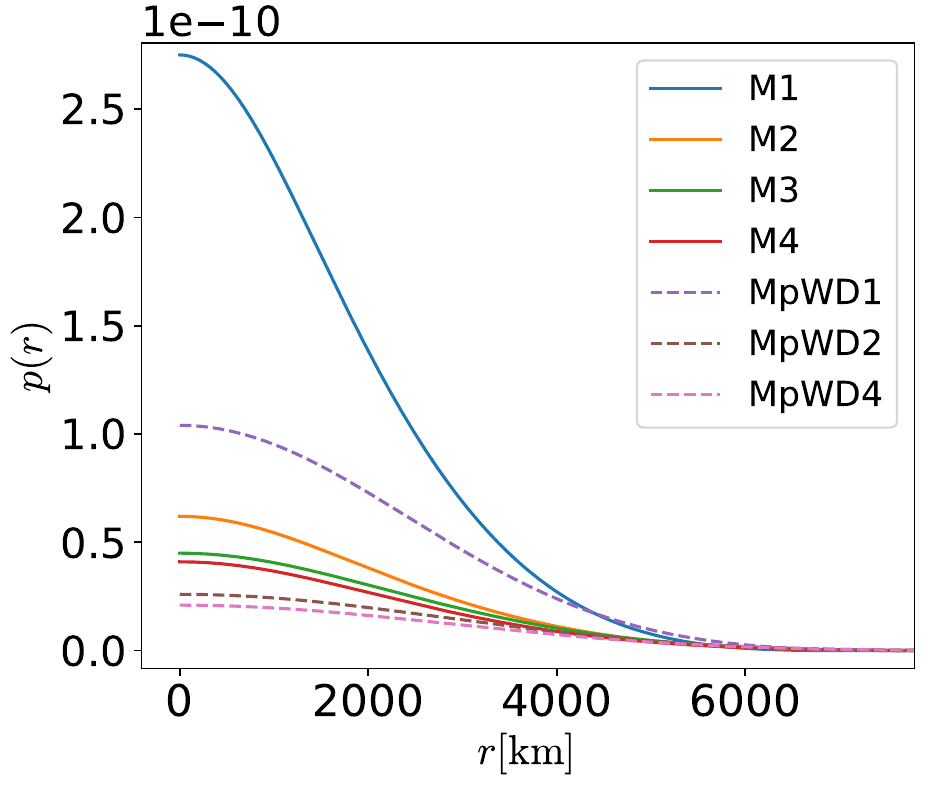}\hfill
    \caption{Top row: scalar field profiles of the models presented in Table~\ref{table1}. Bottom row: pressure profiles for the same configurations. Each column shows the profiles for the different boson masses, as indicated in the titles of the figures. The legends indicate the model names $M_i$, with $i=1,2,3,4$. These numbers correspond to equilibrium models that can be respectively identified with the following observational WDs: 1 for Sirius B, 2 for DB, 3 for DBA, and 4 for Chandra. On the other hand, MpWD refers to pure WD models (with no scalar field component). Only equilibrium solutions with label numbers $1$, $2$, and $4$ are displayed in this case, since the pure WD configurations for cases $2$ and $3$ are identical.
 }
    \label{fig4}
\end{figure*}

Several important conclusions can be drawn from \Cref{table1}. Remarkably, our families of equilibrium configurations successfully reproduce the observational WD data listed in the first column of the table. Both the total mass and the individual contributions from the fermionic (WD) and bosonic (scalar field) components closely match the values inferred from GRS measurements. Furthermore, the fermionic mass component alone corresponds well with the mass estimates obtained through EM methods, thereby reproducing the observed mass difference $\Delta M$ as a natural consequence of the scalar field's contribution. A graphical representation of our equilibrium models is shown in Figure~\ref{fig4}. This figure displays the radial profiles of the scalar field (top row) and of the pressure (bottom row) for all models of \Cref{table1}, highlighting their internal structure and the relative contribution of each matter component. In addition to the total pressure for the mixed models, the bottom panel of Fig.~\ref{fig4} also shows the pressure profile of a purely fermionic WD that would be required to match the EM-based mass estimates. This comparison highlights how the presence of the scalar field significantly modifies the internal structure of the star, allowing the fermionic component to remain consistent with EM observations while the total mass aligns with the GRS-inferred value.

\begin{figure*}
    \centering
    \includegraphics[width=0.45\textwidth]{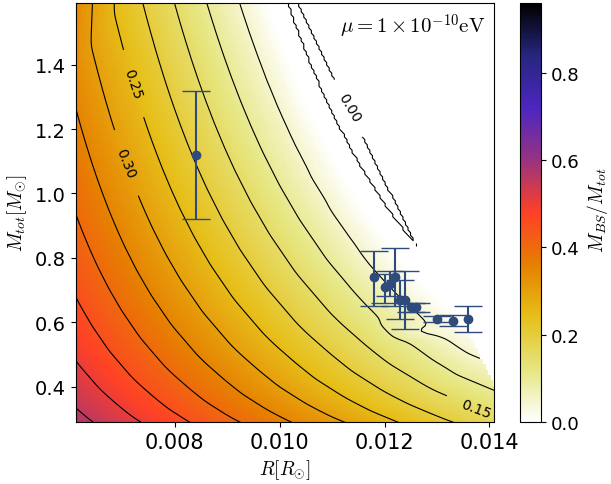}\hfill
    \includegraphics[width=0.48\textwidth]{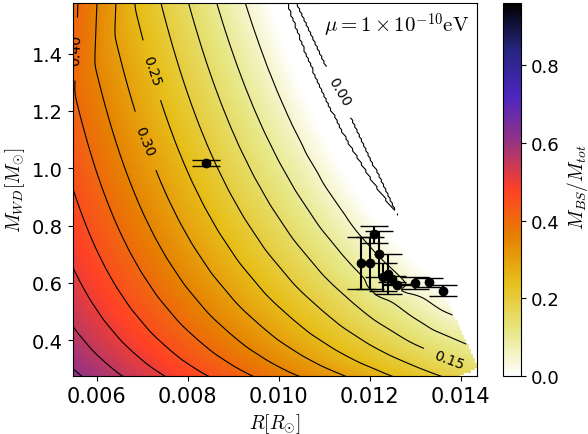}\hfill    
    \includegraphics[width=0.45\textwidth]{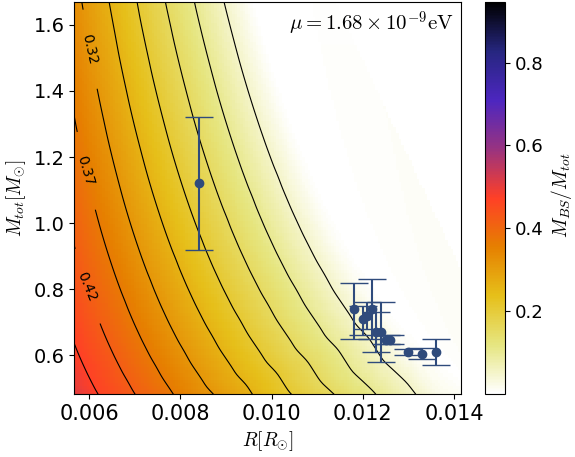}\hfill    
    \includegraphics[width=0.48\textwidth]{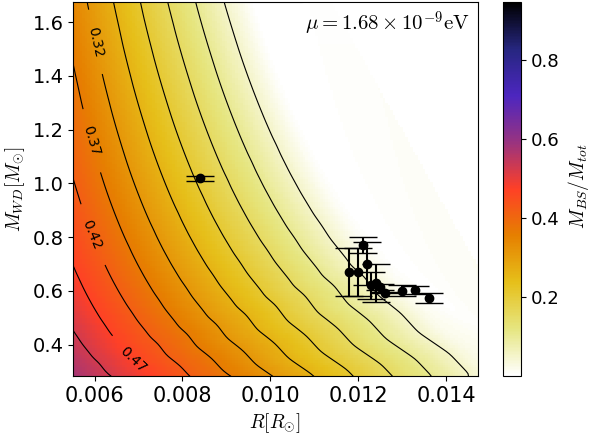}\hfill
    \caption{Comparison between the observational data and the equilibrium configurations for $\mu = 1 \times 10^{-10}$ eV (top) and $\mu = 1.68 \times 10^{-11}$ eV (bottom). The left column displays the GRS mass estimates while the right column shows the EM estimates alongside the fermionic mass components of the mixed configurations. The colormap represents the percentage of bosonic (or DM) content relative to the total mass in each model. This visual representation enables us to evaluate the contribution of the scalar field across various mass regimes and model configurations. The numerical values shown in some of the solid lines indicate the fraction of bosonic matter relative to the total mass, defined as the ratio of bosonic to fermionic-plus-bosonic matter. These values therefore quantify the bosonic contribution associated with each contour line.}
    \label{fig5}
\end{figure*}

The observed WD radii, ranging from approximately $5800\,\mathrm{km}$ ($\sim 0.0084\,R_{\odot}$) to $9000\,\mathrm{km}$ ($\sim 0.013\,R_{\odot}$), are in excellent agreement with the radial scales predicted by our models. An exception arises in the case of the first observational target (Syrius B), where a discrepancy is observed between the mass and the radius. At first glance, it appears that the radius is too small to achieve a good fit. However, a closer examination of the results reveals that the issue lies instead with the mass, which is insufficient to fit within our initial framework. This is particularly visible in Fig.~\ref{fig5}, that we will discuss below. It is worth emphasizing that the equilibrium configurations were obtained by choosing initial parameters, namely, the central pressure and scalar field amplitude, guided by physical intuition and empirical fitting. While a more exhaustive parameter search might yield even closer agreement with observations, the current results already offer a robust and sufficiently accurate description for the purposes of our analysis.

Taken together the results of Table~\ref{table1} and Fig.~\ref{fig4}, our findings support the hypothesis that coupling a scalar field, with bosons of mass $\mu \sim 10^{-11}\,\mathrm{eV}$, to a WD can account for the discrepancies between EM and GRS mass measurements. This mechanism appears viable at least within a well-defined region of the parameter space associated with more massive WDs.

Several remarks emerge when comparing the different models depicted in Fig.~\ref{fig4}. The first column of this figure shows the solutions corresponding to $\mu = 1 \times 10^{-10}\,\mathrm{eV}$. In this case, the scalar field is confined to the innermost region of the star, extending from the center out to only several hundred kilometers, substantially decreasing before reaching $800\,\mathrm{km}$. As expected, the scalar field profile varies depending on the initial amplitude $\phi_0$. The pressure profile, in contrast, spans much larger radial distances, reaching several thousand kilometers and defining total stellar radii between approximately $6000$ and $8500\,\mathrm{km}$, in line with the observed WD sizes. These two types of profiles indicate that the scalar field forms a compact core-like region within the star. This core structure, while limited in spatial extent, contributes significantly to the total mass and has a noticeable impact on the profile of the pressure distribution of the fermionic component. In particular, the scalar field core significantly distorts the pressure profile of a pure WD (see the sharp increase around $r\sim1000$ km in the bottom left panel of Fig.~\ref{fig4}). The required range of initial central pressures is also notably different, varying by at least an order of magnitude from the purely fermionic case. This suggests that, although our models reproduce observational quantities such as total mass, DM fraction, and radius with high fidelity, they may introduce structural differences that could be detectable through other observables. Specifically, the compact scalar field core may influence the WD properties such as the star’s moment of inertia, temperature, or vibrational (asteroseismological) modes. Nonetheless, the model remains consistent with current observations and supports the viability of scalar-core WDs in explaining the EM–GRS mass discrepancy.

The second column of Fig.~\ref{fig4} displays the equilibrium models obtained for $\mu = 5.04 \times 10^{-11}\,\mathrm{eV}$. 
In this case, the scalar field occupies a significantly larger region within the star than for models with $\mu = 1 \times 10^{-10}\,\mathrm{eV}$, extending well beyond $1000\,\mathrm{km}$ before falling off to zero. This more extended scalar distribution has a clear impact on the fermionic pressure profile. Although the pressure remains visibly altered compared to the purely fermionic case, the deviation is less pronounced than in the models with $\mu = 1 \times 10^{-10}\,\mathrm{eV}$. In particular, the pressure curves closely resemble the shape of the pure WD profile, and the initial required central pressures are also more closely aligned. These configurations correspond to scenarios in which the scalar field is distributed across both the inner and intermediate layers of the star. The resulting structural deviation from a standard WD remains significant, but is less extreme than in models where the scalar field forms a compact core. As such, these intermediate cases can offer a compelling balance, allowing for a substantial scalar field contribution while maintaining pressure and density profiles that closely resembles those of conventional WD models.

Finally, the third column of Fig.~\ref{fig4} depicts the equilibrium profiles corresponding to $\mu = 1.68 \times 10^{-11}$ eV. Arguably, this configuration is the most astrophysically plausible, as the scalar field now occupies a radial extent comparable to that of the fermionic matter, rather than being confined to a compact core or limited to intermediate layers. In this regime, the effect of the scalar field on the fermionic component distribution becomes significantly less pronounced. The pressure profile closely resembles that of a pure WD, without extreme deformations. Moreover, the central pressure values required to produce solutions that match the observed masses and radii are of the same order of magnitude to those of a standard WD without a scalar field. As a result, this scenario represents the least disruptive modification to the internal structure of the WD. It provides a natural explanation for the observed discrepancy between GRS and EM mass estimates while preserving the structural and thermodynamical properties expected of a conventional WD. 

\subsection{Models with fixed central scalar field value}

We can further constrain the class of models capable of explaining the observed GRS-EM mass discrepancies by generating sequences of equilibrium solutions with a fixed central value of the scalar field and systematically varying the central fermionic pressure. This procedure allows us to construct families of solutions corresponding to different initial scalar field amplitudes. By comparing these solution sequences with observational mass-radius data, we can identify the configurations that most closely reproduce the observed stellar properties.

Fig.~\ref{fig5} shows existence maps for two values of the mass of the bosonic particle, $\mu = 1 \times 10^{-10}\,\mathrm{eV}$ (top) and $\mu = 1.68 \times 10^{-11}$~eV (bottom). The color scale indicates the percentage of the total mass of the star  contributed by the scalar field. The symbols correspond to the same observational sample of Fig.~\ref{stars_obs}, with their mass uncertainties. For both boson masses, the panels on the left column compare the total equilibrium mass with the GRS-inferred masses. In contrast, the right panels compare the fermionic mass component with the EM-based estimates. This dual comparison enables us to assess whether our models are consistent across the entire parameter space, encompassing both mass and radius, and to visualize the global behavior of the solutions.

Fig.~\ref{fig5} indicates that we can build composite WD models with a fixed central scalar field value that exhibit excellent agreement with observational data for lower-mass WDs, simultaneously reproducing both masses and radii with high accuracy. These models require an amount of scalar field mass contribution to the total mass of the star in the range $10\%-20\%$.
A notable exception arises in the case of our first observational data point, corresponding to Sirius B. In the representative  models discussed in the previous section, we were able to match the mass of Sirius B but the predicted radius (over 6300 km) was larger than the observed value ($5800$ km). In contrast, the solution from the sequence shown in Fig.~\ref{fig5} matching the observed radius of Sirius B corresponds to a configuration with approximately $20\%$ of the total mass in the form of scalar bosonic matter. This discrepancy likely reflects a limitation of the polytropic approximation used to describe the thermodynamics of the fermionic component. In particular, our equilibrium model is not compact enough within our polytropic approximation (its mass is not sufficiently large for the model radius). We address this issue below by providing an alternative fit. However, apart from this single high-mass case, the agreement found between our models and observational data for low-mass WDs is fairly accurate.

By building sequences with a fixed value of the central scalar field we can also gain insight into the plausible mass range of the scalar boson. Because both the fraction of the total mass attributed to the scalar field and the characteristic size of the configuration depend strongly on the boson mass $\mu$, only specific values of $\mu$ yield physically consistent models. In particular, $\mu$ determines the compactness and spatial extent of the scalar field, thereby constraining the regions of parameter space where viable models can exist. As a result, even in the absence of a direct detection, our analysis allows us to infer an approximate order of magnitude for the boson mass. This constitutes an important outcome of this study, as it narrows the viable parameter space for scalar field models in astrophysical contexts.

\renewcommand{\arraystretch}{1.2} 
\begin{table*}[t!]
\renewcommand{\arraystretch}{1.2}
\centering
\caption{Representative parameters for equilibrium models of Sirius B with different scalar field masses. The polytropic parameters are $K=5000$ and $n=1.48$.}
\begin{tabular}{c|cccc|ccccc|cc}
\hline
\textbf{WD} & $M_{\rm GRS}$ & $M_{\rm EM}$ & $R$ & $\Delta M$ 
& $M_{\rm tot}$ & $M_{\rm WD}$  & $M_{\rm SF}$ & $\% M_{\rm bosonic}$  & $R$ & $p_{0}$ & $\phi_{0}$ \\
 & $(\mathrm{M_{\odot}})$ 
 & $(\mathrm{M_{\odot}})$ 
 & $(R_{\odot})$
 & $(\mathrm{M_{\odot}})$ 
 & $(\mathrm{M_{\odot}})$
 & $(\mathrm{M_{\odot}})$
 & $(\mathrm{M_{\odot}})$
 & 
 & (km)
 & (code units)
 & (code units)\\
\hline
\multirow{4}{*}{Sirius B \cite{joyce2018gravitational}}
& \multirow{4}{*}{$1.12$}
& \multirow{4}{*}{$1.018$}
& \multirow{4}{*}{$0.0084$}
& \multirow{4}{*}{$0.102$}
& \multicolumn{7}{c}{$\mu=1\times 10^{-10}\ \mathrm{eV}$} \\
\cline{6-12}
& & & & 
& $1.11$ & $1.01$ & $0.10$ & $\sim 9\%$ & $\sim 5400$ & $84.5\times10^{-10}$ & $1.2\times 10^{-5}$ \\
\cline{6-12}
& & & & 
& \multicolumn{7}{c}{$\mu = 1.68\times 10^{-11}\ \mathrm{eV}$} \\
\cline{6-12}
& & & & 
& $1.11$ & $1.01$ & $0.10$ & $\sim 9\%$ & $\sim 5600$ & $5.5\times10^{-10}$ & $2.2\times 10^{-5}$ \\
\hline
\end{tabular}
\label{table2}
\end{table*}

\subsection{Lower order polytrope}

As shown in Fig.~\ref{fig5}, the first observational point, Sirius B, corresponding to a mass slightly above one solar mass and a scalar matter fraction of approximately $10\%$ (cf.~Table~\ref{table1}), lies in a region of our theoretical models that appears inconsistent. Specifically, the corresponding region of the parameter space suggests a scalar field contribution between $20\%$ and $30\%$, significantly larger than the inferred value. This mismatch can be interpreted as follows.

Our equilibrium models struggle to reproduce the observed radius of this star, consistently predicting a larger value than the measured one. Furthermore, following the contour lines in the scalar mass fraction diagrams reveals that this configuration lies outside the general trends predicted by our model sequences. The underlying reason stems from the simplistic EOS used for the fermionic matter component.
In particular, our model assumes a polytropic EOS with a fixed index, which, while effective for many WD configurations, may introduce some limitations for extreme cases such as this one. Moreover, attempts to fit this star as a pure WD, without a scalar field component, encounter similar difficulties in simultaneously matching the observed mass and radius. This suggests that the issue is not due to the inclusion of the scalar field but is indeed related to the limitations of the polytropic approximation. 

To address this, we explore a slight modification of the polytropic index to improve the model’s flexibility. More specifically, we now take a slightly lower polytropic index, from $n = 1.5$ to $n = 1.48$, while keeping the polytropic constant $K$ fixed. This subtle adjustment allows for solutions that are both less massive and smaller, precisely the kind of structure needed to model Sirius B. With this modified index, in combination with scalar field configurations, we can match the observed mass and radius much more accurately. This is shown in Table~\ref{table2} where we observe that for both the highest and lowest boson mass configurations, slightly reducing the polytropic index leads to equilibrium models that more closely match Sirius B observational data. This adjustment improves the agreement in mass and radius, and consequently yields a more accurate estimate of the scalar field's contribution to the total mass. In particular, the $\mu = 1.68 \times 10^{-11}\rm eV$ case matches the observations with the highest level of agreement between all fitted quantities. While this approach may appear ad hoc, it is grounded in physical reasoning: polytropic models are known to provide valid approximations only within specific regions of the stellar parameter space. Both the index and coefficient are idealizations, and small variations in $n$ are often justified when aiming to extend the validity of the model across a broader class of stars. In this case, the adjustment is minimal but significantly improves the model's ability to capture the observed properties of the most massive WD in our sample.

Regardless of our promising results, it should be emphasized that any description based on a polytropic EOS is necessarily an effective one, whose quantitative predictions depend on the specific EOS adopted, especially when approaching more compact configurations. In the absence of a precisely known EOS for dense fermionic matter, such simplified models remain, however, well motivated and adequate for the goals of the present analysis. At the same time, this framework naturally leaves room for future improvements. In particular, the incorporation of more realistic, possibly temperature-dependent EOS constitutes the main next enhancement of our approach.

\subsection{Mass difference comparison}

\begin{figure}[]
\includegraphics[clip,width=1.0\columnwidth]{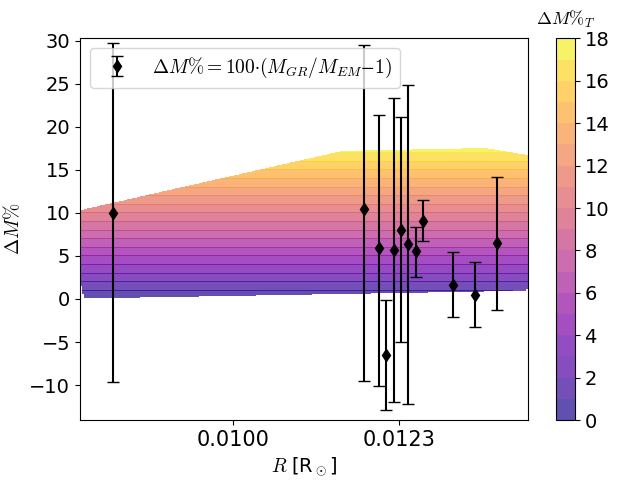}\\%
\includegraphics[clip,width=1.0\columnwidth]{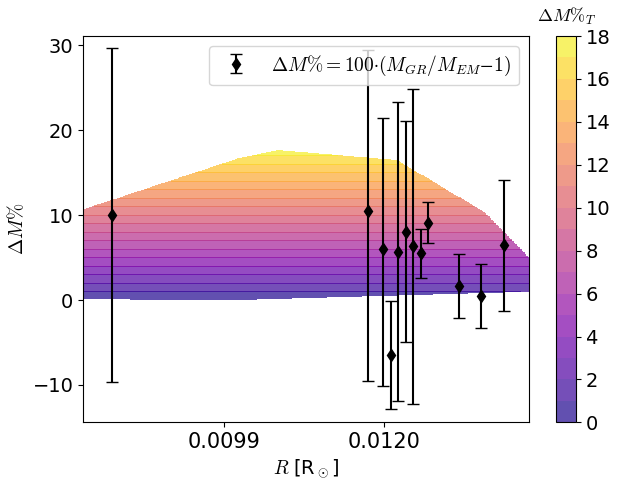}%
\caption{Comparison of the percent mass difference between the observational results ($\Delta M\%$) and our theoretical models ($\Delta M\%_{\rm{T}}$). The top panel corresponds to $\mu = 1 \times 10^{-10}\,\mathrm{eV}$ and the bottom one to $\mu = 1.68\times 10^{-11}\,\rm eV$.}
\label{deltas}
\end{figure}

The comparison of the mass difference between the observational results and our theoretical models is displayed in
Fig.~\ref{deltas} as a function of the stellar radius. The two panels depict the percent of $\Delta M$ for both the observational data (labelled $\Delta M\%$ and indicated by the data points) and its theoretical estimate computed from our equilibrium models (labelled $\Delta M\%_{\rm{T}}$ and shown as the color-shaded region). The error bars associated with each type of mass difference were obtained through error propagation, as explained in \Cref{AppendixB}. The observational mass difference is computed from the difference between GRS and EM estimates while the theoretical one is defined as the difference between total and fermionic masses. The top panel corresponds to $\mu = 1 \times 10^{-10}\,\mathrm{eV}$ and the bottom one to $\mu = 1.68\times 10^{-11}\rm eV$. For the observational data the difference between EM and GRS mass measurements is larger for WDs with smaller radii. The maximum discrepancy observed is $\sim 10\%$. 

On the other hand, the theoretical estimates $\Delta M\%_T$ depicted in Fig.~\ref{deltas} span, for a given radius, a range of possible values. This reflects the fact that different initial scalar field amplitudes yield varying contributions from bosonic DM. This variation is expected and physically reasonable, since the scalar field configuration is sensitive to the initial conditions. We note that the models corresponding to smaller radii in Fig.~\ref{deltas}, particularly those approaching the observed radius of Sirius B, were obtained using the modified polytropic index discussed before. This adjustment proved essential for reproducing the compactness required by the observational data. Overall, our theoretical predictions match the observational measurements of $\Delta M$ with high precision, demonstrating the robustness of our approach across different regions of the parameter space.


\section{Data analysis: model selection}
\label{section7}

\subsection{Statistical framework}

We now quantify the statistical preference of our data for each of our models, characterized by a boson mass $\mu$. Our experimental dataset consists of measurements of the WD radii $R$ together with gravitational and electromagnetic estimates of their masses, denoted by $M_{\rm GRS}$ and $M_{\rm EM}$, respectively. For each star, we consider that the estimates of $M_{\rm GRS}$ and $M_{\rm EM}$ are given by two \textit{independent} Gaussian distributions with means $\hat M_{\rm GRS}$ and $\hat M_{\rm EM}$ and standard deviations $\sigma_{M_{\rm GRS}}$ and $\sigma_{M_{\rm EM}}$. We assume no uncertainty on the radius. That is, for each star $i$  our data is given by
\[
X^i = \{R^i, \hat M_{\rm GRS}^i, \hat M_{\rm EM}^i, \sigma_{M_{\rm GRS}}^i, \sigma_{M_{\rm EM}}^i\}.
\]

For a given $\mu$, our model allows for a range of total masses $M_{\rm tot} \in [M_{\rm tot}^{\rm min}, M_{\rm tot}^{\rm max}]$ for each radius $R$, with an associated unique value of the scalar-field mass $M_{\rm SF}(M_{\rm tot}, R, \mu)$. Thus, the electromagnetic mass predicted by our model is given by
\[
M_{\rm EM}(M_{\rm tot}, R^i, \mu) = M_{\rm tot} - M_{\rm SF}(M_{\rm tot}, R^i, \mu).
\]
With this, the likelihood for a single star is
\begin{widetext}
\begin{equation}
p(X^i|\mu,R^i) =  \int_{M_{\rm tot}^{\rm min}}^{M_{\rm tot}^{\rm max}} p(M_{\rm tot})
\exp\Bigg[-\frac{(\hat M_{\rm GRS}^i - M_{\rm tot})^2}{(\sigma_{M_{\rm GRS}}^i)^2} 
- \frac{(\hat M_{\rm EM}^i - M_{\rm EM}(M_{\rm tot}, R^i, \mu))^2}{(\sigma_{M_{\rm EM}}^i)^2} \Bigg] dM_{\rm tot},
\label{eq:likelihood}
\end{equation}
\end{widetext}
where we will assume a flat prior for $M_{\rm tot}$ given by 
\[
p(M_{\rm tot}|R^i) = \frac{1}{M_{\rm max}(R^i) - M_{\rm min}(R^i)}.
\]
For the full dataset $d = \{X^i\}$, assuming independent stars, the total likelihood is given by the product over stars,
\[
p(d|\mu) = \prod_i p(X^i|\mu),
\]
or, equivalently, the log-likelihood is
\[
\log p(d|\mu) = \sum_i \log p(d|\mu).
\]

We evaluate each single-star likelihood using the \texttt{Dynesty} sampler with 1000 live points. Assuming an uniform prior $p(\mu)$ on $\mu$, the posterior probability is simply given by
\[
p(\mu|d) \propto p(\mu)p(d|\mu) \propto p(d|\mu).
\]

Finally, we also compute the likelihood $p_0 = p(d | M_{\rm GRS} =  M_{\rm EM})$ for a model where the gravitational mass matches the electromagnetic one. That is, we evaluate the likelihood that the observed experimental estimates are equal within the corresponding uncertainties. We obtain this by simply replacing the term $M_{\rm EM}(M_{\rm tot}, R^i, \mu)$ by $M_{\rm tot}$ in the second exponential in Eq. \ref{eq:likelihood}.

\subsection{Results}

\begin{figure}[]
\includegraphics[clip,width=1.0\columnwidth]{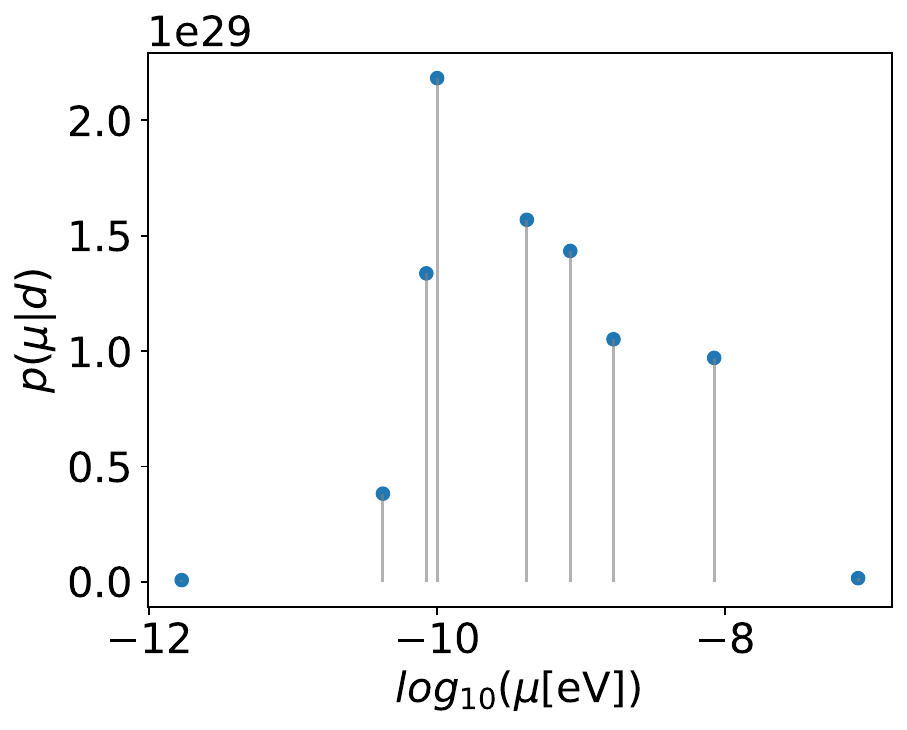}\\%
\caption{Posterior probability for the boson-mass values $\mu$ evaluated considered in this work. We assume a uniform prior probability $p(\mu)$. Thus, the y-axis of the plot is directly proportional to the likelihood $p(d|\mu)$. We have restricted the x-axis to $\mu \geq 1.68\times 10^{-12}$ eV to ease visualization.}
\label{fig:posterior}
\label{deltas2}
\end{figure}

\begin{table}[h]
\centering

\begin{tabular}{|l|c|}
\hline 
$\mu \, [\mathrm{eV}]$ & $p(d|\mu)/p_0$ \\ \hline
$8.41 \times 10^{-17}$ & $0.00$  \\ \hline
$1.68 \times 10^{-12}$ & $0.84$  \\ \hline
$4.20 \times 10^{-11}$ & $37.00$ \\ \hline
$8.41 \times 10^{-11}$ & $ 129.02$ \\ \hline
$1.00 \times 10^{-10}$  & $210.60$ \\ \hline
$4.20 \times 10^{-10}$  & $151.41$ \\ \hline
$8.41 \times 10^{-10}$  & $138.37$ \\ \hline
$1.68 \times 10^{-9}$  & $101.49$ \\ \hline
$8.41 \times 10^{-9}$  & $93.69$ \\ \hline
$8.41 \times 10^{-8}$  & $1.65$ \\ \hline
\end{tabular}
\caption{Likelihood ratios $p(d|\mu)/p_0$ for different values of the boson mass $\mu$ considered in this work.}
\label{tab:ps}
\end{table}

Fig. \ref{fig:posterior} shows the posterior probability distribution $p(\mu|d)$ evaluated at the values of $\mu$ considered in this work. The corresponding numerical values are quoted in Table \ref{tab:ps}, relative to $p_0$. First, we note that the $p_0 = p(d | M_{\rm GRS} =  M_{\rm EM})$ hypothesis is rejected with respect to our scalar-field models with relative probabilities of order $\sim 1/100$ for all values of $\mu$ ranging in $\sim [10^{-11},10^{-9}]$ eV. This means that if we consider two competing models where all these stars either have equal gravitational and electromagnetic masses or they all have contributions to their gravitational mass \textit{sourced by the same boson}, the former hypothesis is rejected with a probability $p \geq 0.99$. While due to the high computational cost of our simulations we cannot reconstruct the full posterior distribution for $\mu$, we can use the discrete set of points as a proxy for it. With this, we can constrain $p(\mu < 1.68 \times 10^{-12}) \leq 1.5\times 10^{-6}$. We note that in this analysis, we do not consider the scenario of ``mixed'' models allowing for a fraction $\zeta_0$ of stars with $M_{\rm GRS} = M_{\rm EM}$ and a fraction $1-\zeta_0$ with scalar-field contributions to $M_{\rm GRS}$. Similarly, we also omit the possibility that several bosons with different masses may contribute to the $M_{\rm GRS}$ value of different stars.\\

Finally, we compute the level of preference of the data for our full model -- spanning all of our mass range for $\mu$ -- and that setting $M_{\rm GRS} = M_{\rm EM}$. We do this by computing the ratio of the marginalized likelihood or Bayesian evidence for our bosonic field model $p_\mu(d) = \int \pi(\mu) p(d|\mu) d\mu$ and $p(d|M_{\rm GRS} = M_{\rm EM})$. With this we obtain 

\[
\frac{p_\mu(d)}{p_0} = \frac{p_\mu(d)}{p(d|M_{\rm GRS}=M_{\rm EM})} \simeq 56.
\]

This is, our mixed-WD model is around 50 times more probable than a model imposing $M_{\rm GRS} = M_{\rm EM}$, thus omitting a scalar field or any other physics that can explain the observed differences between both mass measurements in WDs.\\

\subsection{Physical interpretation}

Here we aim to provide a qualitative physical justification for the above quantitative results. The boson mass $\mu$ is a key parameter that sets the overall scale of the problem. For neutron-star-like configurations, typical static solutions are found within the range $\mu\in[10^{-8},10^{-10}] \,\mathrm{eV}$. The total mass of the system also depends on this parameter, and boson stars with masses close to one solar mass can indeed be obtained within this regime. Conversely, very small values of  $\mu$ correspond to configurations with galactic-scale sizes and halo-like masses \cite{Mourelle:2025ilv}. In our case, we require configurations larger than neutron stars but still relatively compact. This is, we need slightly small $\mu$ values, while keeping the total bosonic mass around $0.1M_\odot$. The crucial point is that there exists a range of $\mu$ values resulting in characteristic lengths between a few kilometers and several thousand kilometers, which are precisely those that yield models consistent with observable scales. Within this region (see Fig.~\ref{deltas2}), larger values of  $p(d|\mu)/p_0$ indicate the $\mu$ range corresponding to the appropriate length scales (cf.~Table~\ref{table2}).\\

While varying $\mu$ values generally lead to different total masses, the field amplitude at the origin $\phi_0$, is an equally important parameter. It allows a broad range of equilibrium configurations, from low-mass to very low-mass models. In general, smaller $\mu$ values produce more extended, less compact configurations, while larger $\mu$  values yield denser, more NS-like stars. Thus, models with higher $\mu$  can still produce a small bosonic mass fraction, if $\phi_0$ is properly tuned, concentrated in a compact core. As $\mu$ decreases, the bosonic distribution expands, the mass becomes more diffuse, and the resulting objects resemble WDs surrounded by excessively large halos. Even if the total mass or $\Delta M$ appear reasonable, such configurations tend to have unrealistically large radii and require increasingly fine-tuned values of $\phi_0$. Finally, the impact of the bosonic dark matter fraction is more pronounced in less massive stars. When the characteristic scales of the WD and bosonic components differ greatly (small $\mu$), the system becomes highly sensitive to small variations in $\mu$. Conversely, for larger $\mu$, variations mainly shift the size of the bosonic core while leaving the WD scale largely unchanged, making these models more stable and physically realistic. We interpret this as the reason behind the fact that Fig. \ref{fig:posterior} shows a very abrupt start of the posterior probability for boson masses around $\mu \simeq 10^{-10}$ eV while it decays very slowly for large masses (note that the x axis is in $\log_{10}$ scale). 

\section{Discussion and conclusions}
\label{section8}

We have put forward a theoretical model to account for the observed discrepancies in the estimation of the mass of white dwarfs derived from electromagnetic methods and through gravitational redshift measurements~\cite{raddi2025testing,arseneau2024measuring,crumpler2024detection}. Our model builds on the work of~\cite{Sanchis-Gual:2022ooi} and relies on the presence of a bosonic scalar field gravitationally coupled to the fermionic component of the star. This dark matter contribution adds up to the total stellar mass yet remains electromagnetically invisible. We have numerically built families of such composite WD solutions showing that those provide a physically motivated explanation for the observed mass bias. These models remain consistent with established astronomical constraints, including the mass–radius relation and luminosity evolution of WDs, even when the bosonic component provides up to $15\%$ of the total mass. Moreover, using our theoretical framework in combination with Bayesian model selection we have also been able to place preliminary constraints on the mass of the constituent (ultralight) bosonic particle. 

The boson masses used in our study fall within the theoretically expected range, and the resulting (macroscopic) configurations reproduce the observed stellar masses and radii with high accuracy. Among the cases studied, the configuration with $\mu = 1.68 \times 10^{-11}\,\mathrm{eV}$ appears particularly compelling, as it leads to fermionic pressure profiles closely resembling those of a pure WD, suggesting minimal structural disruption. In contrast, larger boson masses result in more compact, core-like DM distributions, which alter the internal structure while still fitting global observables. However, after performing a Bayesian analysis with various models spanning different values of $\mu$, we find that the highest likelihood  corresponds to the model with $\mu = 10^{-10}\,\mathrm{eV}$. This configuration features a bosonic DM core of less than $800 \hspace{0.1cm}\mathrm{km}$, which perturbs the pressure and density profiles but still reproduces the observational quantities with a high degree of accuracy. Nevertheless, configurations in which the bosonic matter is more broadly distributed throughout the entire fermionic region yield comparable likelihoods. This demonstrates the flexibility of scalar field models in accommodating a range of WD configurations.

Proving the astrophysical viability of our theoretical model remains an interesting open issue. WDs, as the dense evolutionary end states of low- and intermediate-mass stars, have emerged as compelling astrophysical laboratories for probing the properties of DM, particularly due to their compactness, relative simplicity, and abundance in galactic environments. The possibility that composite stars made out of DM (bosonic or otherwise) and a fermionic fluid might form has been discussed in recent works~\cite{HoefkenZink:2024hor, Bell:2024qmj, Sanchis-Gual:2022ooi,Niu:2024nws}. Crucial and specific to our model, the scalar field DM must have been present during or prior to the WD formation phase, as conventional DM capture over stellar lifetimes yields sub-percent mass fractions~\cite{Dasgupta:2019juq,Graham:2018efk}. Although such a primordial or co-evolutionary origin may seem speculative, it avoids the fine-tuning challenges faced by purely capture-based scenarios and suggests a new class of \textit{dark-core} or \textit{mixed} WDs. 

DM particles traversing a WD may scatter off its constituents: ions, nuclei, or electrons, losing sufficient energy to become gravitationally bound. This capture process is sensitive to both the mass and the velocity distribution of the DM particles. The capture via multi-energy interactions for sub-GeV fermionic DM is a plausible scenario, considering both scalar and vector mediators~\cite{HoefkenZink:2024hor}. Although capture is more efficient at low energies via elastic scattering, there exists a non-negligible window for resonant and deep inelastic scattering at higher energies, involving nucleon resonances such as $N$ and $\Delta$. This highlights the relevance of considering a broad range of DM kinetic energies when modeling WD-DM interactions.

In contrast, as discussed in \cite{Bell:2024qmj}, there is a plausible scenario that shows a heavy DM environment, for which multiple scatterings are necessary before the particle becomes gravitationally captured. The improved treatment carried out by \cite{Bell:2024qmj} accounts for the WD's radial density structure, escape velocity profile, and ion lattice effects. Particularly interesting is their calculation of thermalization timescales, which are found to be significantly shorter if the WD has a crystallized core, due to enhanced in-medium interactions such as phonon emission and absorption. These results refine the parameter space in which asymmetric DM could accumulate and potentially self-gravitate within the stellar core.

Perhaps the most intriguing scenario arises when considering ultralight bosonic DM, as reported in~\cite{Sanchis-Gual:2022ooi}. Using fully relativistic simulations of the Einstein-Klein-Gordon-Euler system, these authors explored the formation of stable mixed configurations composed of a WD and a boson star core. The gravitational interaction between the fluid and the scalar field leads to a dynamical migration of the WD toward a denser and more compact object via a mechanism known as gravitational cooling (also observed in the context of mixed neutron stars~\cite{DiGiovanni:2020frc}). Our results agree with the findings of~\cite{Sanchis-Gual:2022ooi}. Such a structural transformation of the compact star modifies the gravitational redshift and, hence, the observed electromagnetic spectrum. Even in the absence of a direct DM detection, these subtle spectroscopic shifts could serve as observational signatures of the presence of DM in WDs.

Complementing the above approaches, \cite{Niu:2024nws} recently proposed the use of pulsating WDs to constrain DM properties by comparing observed secular variations in pulsation periods with those predicted by stellar evolution models. While this method currently provides less stringent bounds, it remains a promising avenue to indirectly assess DM capture, annihilation, and evaporation, particularly in scenarios involving DM-electron interactions. Additional channels are those that connect the measurement of WDs in binary systems with GW observations. This methodology, described and analyzed in \cite{Sala:2025uqh}, could provide valuable insights into the potential DM content of these objects.

This discussion shows that the landscape of DM interactions in WDs is rich with theoretical and observational possibilities. Among the various candidates, ultralight bosonic fields forming hybrid stars stand out as particularly transformative, since they provide a self-consistent relativistic framework and predict astrophysical signatures such as shifts in mass–radius estimates or spectral redshifts. High-precision WD surveys and asteroseismology may therefore offer a uniquely sensitive probe of DM and its couplings to baryonic matter. Additional insights can be gained in future observational campaings including microlensing events, astrometric binaries, and more precise redshift measurements. These could help identify candidates with substantial non-baryonic components and offer new tests of exotic matter contributions in compact stars.

The main limitation of this study arises from the scarce observational sample of WDs with independently available GRS and EM measurements, which restricts the scope of the analysis. To mitigate selection bias, all objects for which both measurements could be identified were included, without any \emph{a priori} filtering. Although the observed trends may suggest a common underlying behavior, the limited sample size precludes any claim of strict universality. Instead, our results should be interpreted as evidence for a quasi-systematic difference that emerges in those systems where direct GRS and EM comparisons are feasible.

Several improvements to our model can be envisaged. One obvious possibility is to refine the treatment of the fermionic EOS, moving beyond the polytropic approximation to incorporate tabulated equations of state, finite-temperature effects, and composition-specific parameters. Moreover, exploring other bosonic sectors, such as vector fields, may also extend the theoretical landscape. Results from those future studies will be reported elsewhere.

 \begin{acknowledgements}
JCM thanks Rocío Bello Mallo for her inspiration on this topic. JCM and NSG acknowledge support from the Spanish Ministry of Science, Innovation and Universities via the Ram\'on y Cajal programme (grant RYC2022-037424-I), funded by MCIN/AEI/10.13039/501100011033 and by ``ESF Investing in your future”. JCM is also funded by a UNAM-DGAPA Postdoctoral Fellowship. NSG and JAF are supported by the Spanish Agencia Estatal de Investigaci\'on (grants PID2021-125485NB-C21 and PID2024-159689NB-C21) funded by MCIN/AEI/10.13039/501100011033 and ERDF A way of making Europe, and by the Generalitat Valenciana (grant CIPROM/2022/49). JCB is supported by the Ramon y Cajal Fellowship RYC2022-036203-I and the Grant PID2024-160643NB-I00 of the Spanish Ministry of Science, Innovation and Universities; and by the Grant ED431F 2025/04 of the Galician Conselleria de Educaci\'on, Ciencia, Universidades e Formaci\'on Profesional. IGFAE is supported by the Ayuda Maria de Maeztu CEX2023-001318-M funded by MICIU/AEI/10.13039/501100011033. We acknowledge further support from the European Horizon Europe staff exchange (SE) programme HORIZON-MSCA2021-SE-01 Grant No. NewFunFiCO-101086251 and by the
Center for Research and Development in Mathematics and Applications (CIDMA) through
the Portuguese Foundation for Science and Technology (FCT – Fundação para a Ciência
e a Tecnologia) under the Multi-Annual Financing Program for R\&D Units,  2024.05617.CERN (\url{https://doi.org/10.54499/2024.05617.CERN}). 
\end{acknowledgements}


\appendix
\section{Units and transformations}
\label{AppendixA}

In this appendix we explain the internal units used for numerically solving our models, as well as the changes needed to recover physical units.

\subsubsection{Unit system}

We adopt a geometrized unit system where $G = c = 1 $, but keep explicit physical units for conversion to astrophysical scales. Some key constants are:

\begin{itemize}
  \item \( \hbar c = 197.327 \, \mathrm{MeV \cdot fm} \)
  \item \( 1 \, \mathrm{fm} = 10^{-13} \, \mathrm{cm} \)
  \item \( 1 \, \mathrm{MeV/fm}^3 = 1.7827 \times 10^{12} \, \mathrm{g/cm^3} \)
  \item \( M_\odot = 1.9885 \times 10^{33} \, \mathrm{g} \)
  \item \( 1 \, \mathrm{km} = 10^5 \, \mathrm{cm} \)
\end{itemize}

Quantities are internally expressed in combinations of  $\mathrm{MeV}$, $\mathrm{fm}$, $M_\odot$, and  $\mathrm{km}$, to yield physical observables in astrophysically meaningful units.

\subsubsection{Radial scale of the bosonic field}

The mass of the scalar particle  $\mu$  determines the characteristic length scale of the bosonic configuration:
\begin{equation}
  R_{\text{scale}} = \frac{\hbar c}{\mu} \cdot \left( \frac{\mathrm{fm}}{\mathrm{km}} \right)  \,.
\end{equation}

For WD-like configurations, the right value for the scalar field mass is centered around $\mu\sim 10^{-11} \mathrm{eV}$, which allows us to define a radial scale ${\cal R} \sim C_t \hbar c / \mu $ where some dimensional transformations are absorbed in $C_t$. Since $\mu$ is small, the radial extent of the bosonic component can be large (hundreds to thousands of $\rm km$).

\subsubsection{Scalar field normalization}

The scalar field  $\phi_s$  is normalized to have units of $\sqrt{M_\odot/\mathrm{km}}$, as required for compatibility with  Einstein's field equations. The conversion is:
\begin{equation}
   \phi_s = \phi_{\text{code}} \cdot \frac{1}{\sqrt{8\pi \kappa}}, \quad \text{where } \kappa = \frac{G}{c^4} 
\end{equation}

\subsubsection{Pressure and density scales}

The initial pressure of the fermionic component is given in units of $ M_\odot / \mathrm{km}^3 $. The pressure is related to the density through a polytropic EOS, $ P = K \rho^\gamma$ with $ \gamma = 1 + \frac{1}{n}$ and $ n=1.5 $.
For obtaining the numerical models we take the value of the polytropic constant $K$ to be one, as it rescales all quantities.  Physical observables can be obtained afterwards by taking a realistic value for our purpose, which is $K=5000$.

\subsubsection{Boson field frequency scale}

The internal frequency $w_s$ of the bosonic field has units of inverse length. Its scale is
\begin{equation}
   w_{s} = w_{\rm shot} \cdot \frac{1}{{\cal R}} .
\end{equation}
The scalar field potential is defined as,
\begin{equation}
    V(\phi) = \mu^2 \phi^2,
\end{equation}
which is the non-self-interacting (or the so-called mini boson star) potential. This potential is evaluated in units of  $M_\odot / \mathrm{km}^3$, which is consistent with the stress-energy tensor.

\subsubsection{Physical scaling of mass and radius}

To convert dimensionless output to physical units we use a scaling based on the polytropic constant $K$, as mentioned. For the mass and radius, which are the main observables discussed in this work, the transformations read,
\begin{equation}
    \begin{split}
       &M_{\text{phys}} = \frac{M_{\text{code}}}{K^{-n/2}}\,, \\
       &R_{\text{phys}} = \frac{R_{\text{code}}}{K^{-n/2}} \,.
    \end{split}
\end{equation}
This allows proper comparison with observational data (expressed in, e.g., $M_\odot$, km). 

\section{Error propagation in the mass difference estimate}
\label{AppendixB}


In this appendix we explain how the errors associated with the mass estimate, $\Delta M = M_{\rm GRS} - M_{\rm EM}$, are estimated. These estimates are based on available observational data and their reported uncertainties, assuming that the deviations follow a Gaussian (normal) distribution. This approach allows us to determine whether the predicted variations in our models remain consistent with observations within the expected error margins \cite{lindegren2021gaia}.
We study the percent difference of the mass estimate,
\begin{equation}
  \mathrm{pct\_diff}
  = \Bigl(\frac{M_{\rm GRS}}{M_{\rm EM}} - 1\Bigr)\times100\%\,,
\end{equation}
where $M_{\rm GRS}$ and $M_{\rm EM}$ have independent, asymmetric uncertainties
\begin{equation}
    M_{\rm GRS}\pm(\sigma_x^-,\sigma_x^+),
  \quad
  M_{\rm EM}\pm(\sigma_y^-,\sigma_y^+). 
\end{equation}
We define the ratio between masses
\begin{equation}
     r = \frac{M_{\rm GRS}}{M_{\rm EM}},
\end{equation}
with partial derivatives
\begin{equation}
    \frac{\partial r}{\partial M_{\rm GRS}} = \frac{1}{M_{\rm EM}},
  \quad
  \frac{\partial r}{\partial M_{\rm EM}} = -\,\frac{M_{\rm GRS}}{M_{\rm EM}^2}. 
\end{equation}
If errors were symmetric ($\sigma_x^\pm = \sigma_x$, $\sigma_y^\pm = \sigma_y$), then
\begin{equation}
     \sigma_r
  = \sqrt{
    \Bigl(\tfrac{\sigma_x}{M_{\rm EM}}\Bigr)^2
    + \Bigl(\tfrac{M_{\rm GRS}\,\sigma_y}{M_{\rm EM}^2}\Bigr)^2
  },
  \quad
  \sigma_{\rm pct} = 100\,\sigma_r.
\end{equation}
For asymmetric errors one treats “$+$” and “$-$” separately:
\begin{align}
  \sigma_r^+ &= \sqrt{
    \frac{(\sigma_x^+)^2}{M_{\rm EM}^2}
    + \frac{M_{\rm GRS}^2\,(\sigma_y^-)^2}{M_{\rm EM}^4}
  }, \label{eq:sigmaR+}\\
  \sigma_r^- &= \sqrt{
    \frac{(\sigma_x^-)^2}{M_{\rm EM}^2}
    + \frac{M_{\rm GRS}^2\,(\sigma_y^+)^2}{M_{\rm EM}^4}
  }.\label{eq:sigmaR-}
\end{align}
Since $\mathrm{pct\_diff}=(r-1)\times100\%$, the percentage errors become
\begin{equation}
    \sigma_{\rm pct}^+ = 100\,\sigma_r^+,
  \quad
  \sigma_{\rm pct}^- = 100\,\sigma_r^-,
\end{equation}
or explicitly
\begin{equation}
  \begin{aligned}
    \sigma_{\rm pct}^+ &= 100\,
    \sqrt{
      \frac{(\sigma_{M_{\rm GRS}}^+)^2}{M_{\rm EM}^2}
      + \frac{M_{\rm GRS}^2\,(\sigma_{M_{\rm EM}}^-)^2}{M_{\rm EM}^4}
    },\\
    \sigma_{\rm pct}^- &= 100\,
    \sqrt{
      \frac{(\sigma_{M_{\rm GRS}}^-)^2}{M_{\rm EM}^2}
      + \frac{M_{\rm GRS}^2\,(\sigma_{M_{\rm EM}}^+)^2}{M_{\rm EM}^4}
    }.
  \end{aligned}
\end{equation}
These expressions are used to compute the error bars displayed in Fig.~\ref{stars_obs}.

It is important to emphasize that this represents an approximate estimate of the error and should not be interpreted as a strict bound. The associated error bars are qualitative in nature, as they arise from underlying assumptions and approximations made in our modeling framework.

\bibliography{biblio}

\end{document}